\documentclass[a4paper,11pt]{article}
\usepackage{jcappub} 
\usepackage[T1]{fontenc}
\newcommand{\be}{\begin{equation}}
\newcommand{\ee}{\end{equation}}
\newcommand{\bea}{\begin{eqnarray}}
\newcommand{\eea}{\end{eqnarray}}

\newcommand{\apj}{{Astrophys. J.}}
\usepackage{multirow}

\title{\boldmath Fundamental oscillation modes of self-interacting bosonic dark stars}

\author[a]{C. V\'{a}squez Flores,}
\author[b]{Alessandro Parisi,}
\author[b]{Chian-Shu Chen,}
\author[c]{Germ\'{a}n Lugones,}

\affiliation[a]{Departamento de Fisica, \rm{CFM}, Universidade Federal de Santa Catarina, Florian\'{o}polis - \rm{SC} - \rm{CP}. 476, \rm{CEP} 88.040 - 900, Brazil}
\affiliation[b]{Department of Physics, Tamkang University, New Taipei 251, Taiwan}
\affiliation[c]{Universidade Federal \rm{do} \rm{ABC}, Centro de Ci\^{e}ncias Naturais e Humanas, Avenida dos Estados, 5001-Bang\'{u}, CEP 09210-170, Santo Andr\'{e}, SP, Brazil}

\emailAdd{cesarovfsky@gmail.com}
\emailAdd{alessandro26@live.it}
\emailAdd{150820@mail.tku.edu.tw}
\emailAdd{german.lugones@ufabc.edu.br}

\abstract{
We perform a detailed analysis of the fundamental $f$-mode frequencies and damping times of nonrotating boson stars
in general relativity by solving the nonradial perturbation equations.
Two parameters which govern the microscopic properties of the bosonic condensates, namely the self-coupling strength and the mass of scalar particle, are explored.
These two quantities characterize oscillations of boson star. 
Specifically, we reexamine some empirical relations that describe the $f$-mode parameters in terms of mass and radius of the boson stars. 
We found it is possible to constrain the equation of state if the fundamental oscillation mode is observed. 
}

\begin{document}
\maketitle
\flushbottom

\section{Introduction}
\label{sec:intro}
Gravitational waves (GWs) from a binary black hole (BH) were detected at the Advanced LIGO interferometer in september 2015~\cite{Abbott:2016blz}. 
It will open a new window to explore the Universe. 
Interestingly the LIGO and Virgo Collaborations reported the first event, GW170817, where a gravitational-wave signal was observed from a merger of two neutron stars (NSs)~\cite{TheLIGOScientific:2017qsa}. Based on the GW170817 observation, several recent studies have reported their constrains on NS equations of state (EoS)~\cite{Ruiz:2017due,Annala:2017llu,Bauswein:2017vtn,Margalit:2017dij,Fattoyev:2017jql,Rezzolla:2017aly,Drago:2017bnf,Nandi:2017rhy}.\\
While BHs and NSs now represent the standard model of compact objects, it is worth exploring alternatives which differ in their GW signatures from the standard one.
In this work, we compute the $f$-mode of an important class of hypothetical objects,
composed of self-interacting scalar field configurations known as boson stars (BSs).
The nature of these objects depends of the scalar self-interaction and its coupling to gravity. 
Examples of such nonstandard stars were widely discussed in literature, such as a geon, which is a self-gravitating star consisting of electromagnetic fields and was first considered by Wheeler~\cite{Wheeler:1955zz}.
The gravitational attraction by its own field energy confines the geon in a certain region. Later Kaup solved the Einstein-Klein-Gordon (EKG) equations for a massive complex scalar field and found a new class of solutions for compact objects~\cite{Kaup:1968zz}.
These BSs are stable with respect to spherically symmetric gravitational collapse.
Ruffini and Bonazzola~\cite{Ruffini:1969qy} demonstrated that BSs describe a family of self-gravitational scalar field configurations within general relativity. 
Although the existence of these elementary bosons, their clustering, and their hypothetical role in the formation 
of galaxies and large scale structure of the Universe, is nowadays obscure~\cite{Liddle:1993ha},
there are two theoretical arguments that supports the possibility of self-gravitating objects made by bosonic particles in the Universe.
First, the discovery of Higgs boson~\cite{Aad:2012tfa,Chatrchyan:2012xdj} confirmed the existence of scalar fields in nature.
Second, the existence of a formation mechanism, dubbed as gravitational cooling~\cite{Seidel:1993zk}, to produce BSs from a generic scalar field configuration.
In the past few years BSs have been studied in many different contexts (see ~\cite{Liebling:2012fv,Jetzer:1991jr,Schunck:2003kk} for  complete reviews).
The stability of BSs  against radial perturbations around the equilibrium state also has been studied by several authors~\cite{Lee:1988av,Jetzer:1988vr,Gleiser:1988rq}.
It is noted that BSs and NSs share a remarkable similarity on the stability properties.  
Furthermore, BS is also considered extensively to be form by dark matter (DM) candidates~\cite{RindlerDaller:2011kx,UrenaLopez:2010ur} 
and alternatives to black holes~\cite{Yuan:2004sv}. 
Torres et al.~\cite{Torres:2002td,Torres:2000dw} showed that BS (for a large range of boson masses and self-interactions) can also constitute a viable alternative for the central supermassive object in the Galactic center.  
Recently Olivares et al.~\cite{Olivares:2018abq} found that the absence of an event horizon in a supermassive BS 
leads to important differences in the dynamics of the accretion flow with respect to the case of a Kerr BH,
and conclude that it is possible to discriminate between a BH and a BS from the accretion process.\\
The gravitational wave production from the merger of two BSs has been studied by Palenzuela et al.~\cite{Palenzuela:2007dm,Palenzuela:2017kcg}
and, in particular, the emission of gravitational radiation by an oscillating BS has been studied under the Newtonian approximation by Ferrell and Gleiser~\cite{Ferrell:1989kz}. It has been shown that the amount of gravitational energy corresponds to 
the transition energy from an excited state to the ground state of the oscillation modes. 
Quasinormal modes of BSs were also obtained within general relativity by Yoshida et al.~\cite{Yoshida:1994xi},
Balakrishna et al.~\cite{Balakrishna:1997ej}, and more recently by Macedo et al.~\cite{Macedo:2013jja}.
Kling and Rajaraman~\cite{Kling:2017mif,Kling:2017hjm} found a semianalytic solution describing dilute BSs in the Newtonian limit,
and showed that the solution is stable to numerical errors.\\
Tidal deformability of BSs has also been investigated in~\cite{Sennett:2017etc, Mendes:2016vdr} 
and this can be used to discriminate between BSs and NSs with the future aLIGO sensitivity.\\ 
In this work, we study the $f$-mode of BS with different self-interaction strengths and different masses of scalar boson.  
Recently these modes have been studied for many models of compact stars in~\cite{Flores:2017hpb,Flores:2018pnn} 
and in the context of binary systems in~\cite{Parisi:2017kgx, Yang:2018bzx}.
The scope of this work is to connect the microscopic properties of scalar boson with the macroscopic observables of BS.
In particular we compute both real and imaginary parts of 
the oscillation frequencies which may be observed if pulsations are excited during the formation of BSs or during their evolution under the action of external perturbation.
The paper is organized as follows. In Sec.~\ref{sec:modes} we present the main properties of quasi-normal modes of compact stars.
Equilibrium configurations and the EoS of the BS are studied in Sec.~\ref{sec:boson}. Sec.~\ref{sec:Results_BS} 
present the $f$-mode with various possible parameters of the EoS. 
We draw our conclusions and the possible connections to astronomical observations in Sec.~\ref{sec:conclusions}.

\section{The quasi-normal modes of compact stars}
\label{sec:modes} 
The equations which describe the nonradial pulsations of a compact star (CS) in a fully general relativistic context were first studied by
Thorne and Campolattaro~\cite{1967ApJ...149..591T,1970ApJ...159..847C}. They showed that Einstein's equations describing small, nonradial, quasi-periodic oscillations of general relativistic stellar models could be reduced to a system of ordinary differential equations for the perturbed functions.
We use the formulation of Lindblom and Detweiler (see Appendix \ref{sec:appA})~\cite{Lindblom:1983ps,Detweiler:1985zz}, where 
Thorne's equations are reduced to a system of four ordinary differential equations and integrated the perturbation equations directly in a manner similar to Thorne. 
These equations describe the fluid oscillations of the star as well as the emitted gravitational waves, thus leading to a damping of the star's oscillation.\\ 
We assume the unperturbed spherically symmetric equilibrium state of a CS is given by a solution of the Tolman-Oppenheimer-Volkhoff (TOV) equations. 
For pulsations of spherical-harmonic indices $\ell$ and $m$ and parity $\pi=(-1)^\ell$, the  perturbed metric
tensor inside the star in the Regge-Wheeler
gauge~\cite{Regge:1957td} is given by
\begin{eqnarray}
  ds^2 &=& -e^\psi(1+r^\ell H_0^{\ell m}Y_{\ell m}e^{i\omega t})dt^2 +e^\lambda(1-r^\ell H_2^{\ell m}Y_{\ell m}e^{i\omega t})dr^2 \nonumber\\
       & &  -2i\omega r^{\ell+1}H_1^{\ell m}Y_{\ell m}e^{i\omega t}dtdr + r^2(1-r^\ell K^{\ell m}Y_{\ell m}e^{i\omega t})(d\theta^2+\sin^2\theta d\varphi^2),
\end{eqnarray}
where $\omega$ is the frequency, $Y_{\ell m}$ denote the usual scalar spherical harmonics, the functions $e^\psi$ and $e^\lambda$
are the components of the metric of the unperturbed stellar model, while $H_i^{\ell m}(r)$ and $K^{\ell m}(r)$ characterize the metric perturbations. 
In this paper we do not consider perturbations with axial parity because they are
not characterized by pulsations which emit gravitational waves~\cite{1967ApJ...149..591T}.\\
The perturbation of the CS fluid is described by the Lagrangian displacement vector $\xi_a$, having components
\begin{eqnarray}
  \xi_r(t,r,\theta,\varphi)        &=& e^{\lambda/2}r^{\ell-1}W^{\ell m}(r)Y_{\ell m}(\theta,\varphi)e^{i\omega t}, \nonumber\\
  \xi_\theta(t,r,\theta,\varphi)   &= &  -r^\ell V^{\ell m}(r)\partial_\theta Y_{\ell m}(\theta,\varphi)e^{i\omega t},\\
  \xi_\varphi(t,r,\theta,\varphi)  &=&   -r^\ell V^{\ell m}(r)\partial_\varphi Y_{\ell m}(\theta,\varphi)e^{i\omega t}.\nonumber
\end{eqnarray}
In this paper we use the formulation of Lindblom and Detweiler~\cite{Lindblom:1983ps,Detweiler:1985zz}, 
consisting of a system of four ordinary differential differential equations 
\be\frac{d\mathbf{Y}(r)}{dr}=
\mathbf{Q}(r,\ell,\omega)\mathbf{Y}(r)\ee 
for the functions $\mathbf{Y}(r)=(H_1^{\ell m},K^{\ell m},{W}^{\ell m},X^{\ell m})$,
where \be X^{\ell m}=-e^{\psi/2}\Delta p^{\ell m} \ee 
and three algebraic relations which allow to compute the remaining functions $\{ H_0^{\ell m},H_2^{\ell m},V^{\ell m}\}$ 
in terms of the others, see Appendix \ref{sec:appA}.
We concentrate our attention on normal modes which belong to a particular even parity spherical harmonic $\pi=(-1)^\ell$ 
with the complex frequency 
\be \omega=\sigma+\frac{i}{\tau} \;.\ee 
The normal modes of the coupled system are defined as those oscillations which lead to purely outgoing waves at spatial infinity. 
The real parts of their eigenfrequencies correspond to the oscillation rate; the imaginary parts describe the damping due to radiative energy loss.\\
A CS at the end of its evolution is cold and isentropic, and can be described by a barotropic EoS $p=p(\varepsilon)$. 
In contrast, in a hot CS the situation is more complicated because the pressure depends nontrivially on entropy $s$, i.e.,
\be p=p(\varepsilon,s). \ee
The thermal effects on a CS have been studied for different oscillation's mode in Burgio et al.~\cite{Burgio:2011qe},
in general the frequencies and the damping times can change a lot with the temperature.
The thermal effects for a self-gravitating BS have been studied in~\cite{Bilic:2000ef,Latifah:2014ima}, 
these effects are very difficult to study because is necessary to use the effective field theory at finite temperature, 
then in general the concept of entropy per particle can also be introduced for a star made up of bosons, 
see the book of Pitaevskii and Stringari~\cite{Gross} for a complete description of the bosonic properties. 
In ~\cite{Latifah:2014ima} the authors find that the EoS for a BS not depend sensitively on the temperature variation,
so that the maximum mass predictions are not significantly different, while at low densities, 
however, the EoS is quite sensitive to the temperature of boson matter.
Unfortunately we do not know  the thermal evolution of the BS  and therefore 
in this paper we prefer to assume adiabatic oscillations for which the BS is described by a barotropic EoS $p=p(\varepsilon)$.

\section{Equilibrium Configurations of the Boson Stars}
\label{sec:boson} 
A BS is a stellar object made of bosons, contrary to conventional stars which are formed of fermions~\cite{Shapiro:1983du}.
They are similar in many respects to NSs, differing in that their pressure support derives from the Heisenberg uncertainty relation rather than the exclusion principle.
The existence of BSs was first theoretically demonstrated by Ruffini and Bonazzola~\cite{Ruffini:1969qy} for a non-interacting case.
They analyzed only the zero-node solutions, corresponding to the lowest energy state. It has been shown that boson stars are stable to small radial perturbations, provided that their central density does not exceed a critical value which also 
corresponds to the configuration with the maximum possible mass~\cite{Gleiser:1988ih}. 
BSs are described by the EKG equations deriving from the action
\be\label{action}  S=\int d^4x  \sqrt{-g}\left[\frac{R}{16\pi G}- \nabla^{\alpha}\Phi \nabla_{\alpha}\Phi^{\ast}-V(|\Phi|^2)\right],\ee  
where $R$ is the Ricci scalar, $\Phi$ is the scalar field, $\Phi^\ast$ its complex coniugate, and $V(|\Phi|^2)$ is the potential.   
Different BS models are classified according to their scalar potential and particle properties.
Here we focus on the self-interaction potential
\be  V(|\Phi|)=\frac{1}{2}m^2|\Phi|^2+\frac{\lambda}{4}|\Phi|^4    \ee
where $m$ is the mass of the field and $\lambda$ its self-interaction.
The presence of a self-interaction in the scalar potential is known to have significant effects on the structure of the BS see Colpi at al.~\cite{Colpi:1986ye}.
The dimensionless ratio $\lambda \Phi^4/m^2\Phi^2$ characterize the relative contribution of the potential energy due to self-interaction to the mass term. 
In the equilibrium state, the BS mass is characterized by $\sqrt{\Lambda}M^2_{\rm{Pl}}/m$ where $M_{\rm{Pl}}$ is the Planck mass and the dimensionless quantity $\Lambda\equiv\lambda M^2_{\rm{Pl}}/(4\pi m^2)$. 
It shows the interaction can dominate the potential energy even for a very small $\lambda$ and mass of BSs can be comparable to the mass of NSs.\\ 
The  action (\ref{action}) is invariant under the $U(1)$ global transformation $\Phi\rightarrow e^{i\theta}\Phi$, we obtain the continuity equation
\be   \frac{1}{\sqrt{-g}}(\sqrt{-g}J^\mu)_{,\mu}=0,   \ee
where the comma denotes differentiation with respect to the following quantity and $J_{\mu}$ is the conserved four-vector current defined by
\be  J_{\mu} = i( \Phi^\ast \nabla_{\mu}\Phi -\Phi \nabla_{\mu}\Phi^\ast).  \ee
The associated Noether charge,
\be Q = \int g^{0\mu}J_{\mu}  \sqrt{-g}\; d^3x,   \ee
can be identified as the boson number.
Note that the conservation of boson number here is due to the complex nature of the scalar; for a real scalar field there is no such conserved charge.
By varying  the action with respect to $\Phi^\ast$ and  $g^{\mu\nu}$, we obtain the scalar field equation 
\be   \nabla^\mu\nabla_\mu\Phi = \frac{dV}{d|\Phi|^2}\Phi.    \ee
and the Einstein equations derived from the action (\ref{action}) are given by
\be   R_{\alpha\beta}-\frac{1}{2}g_{\alpha\beta}R=8\pi T_{\alpha\beta}^{\Phi},     \ee
where $T_{\alpha\beta}^{\Phi}$ is stress-energy tensor of the scalar
\be T_{\alpha\beta}^{\Phi}= \nabla_{\alpha}\Phi^{\ast} \nabla_{\beta}\Phi + \nabla_{\beta}\Phi^{\ast} \nabla_{\alpha}\Phi 
-g_{\alpha\beta}(\nabla^{\gamma}\Phi^{\ast}\nabla_{\gamma}\Phi  +V(|\Phi|^2)).    \ee
We seek for the spherically symmetric solution with a static metric, and the metric can be chosen in the form
\be  ds^2 = -B(r)dt^2+A(r)dr^2+r^2d\theta^2+r^2 \sin^2\theta d\psi^2.   \ee
We assume the field with a time dependence $\Phi(r,t)=\Phi_0(r)e^{-i\omega t}$, and the stress energy tensor is time independent which implies the space-time is stationary and the metric functions depend only on the radial coordinate $r$.
Practically, $\Phi(r,t)$ is a complex field but we can choose our field definition such that the complex part vanishes at time $t=0$.\\
The EKG equations reduce to a system of ordinary differential equations for the metric functions $A$, $B$, and for the scalar field $\Phi$.
The equilibrium configurations are found by numerically integrating the EKG along with suitable boundary conditions.
For a given value $\phi_c$ of the scalar at the center of the star, the problem is then reduced to an eigenvalue problem for the frequency $\omega$.
Colpi et al.~\cite{Colpi:1986ye} showed that, in the Thomas-Fermi limit, corresponding to $\Lambda\gg 1$, 
the scalar field becomes equivalent to a fluid with an EoS
\be\label{press_BS} p= \frac{c^4}{36K}\left[\left(1+\frac{12K}{c^2}\rho\right)^{1/2}-1\right]^2  \ee
with
\be K\equiv \frac{\lambda\hbar^3}{4m^4c}, \ee
where $p$ and $\rho$ represent the pressure and density respectively.
Chavanis and Harko~\cite{Chavanis:2011cz} shows the accuracy of the hydrodynamical approach in this limit.
Therefore under these conditions the BS can be treated as a perfect fluid~\cite{AmaroSeoane:2010qx}. 
In this limit the anisotropy parameter $\delta\equiv(p_r-p_{\bot})/p_r$ where $p_r$ and $p_{\bot}$ 
are the radial and tangential pressure approaches to zero.
The parameter $\delta$ measures the deviation from local isotropy and was investigated by Gleiser~\cite{Gleiser:1988rq} 
with the surprising conclusion that the value of $\delta$ at the surface of the star is only weakly dependent on its central density. 
This EoS was used in Maselli et al.~\cite{Maselli:2017vfi} to study the I-Love-Q universal relations for a BS,
showing that these relations for both the fermion and the boson case do exist, 
and could be extremely useful in the near future to combine multiple observations and perform redundancy tests of the stellar model. 
In the Newtonian limit Eq.(\ref{press_BS}) is in the form of a polytropic EoS with $n=1$,
 \be  p=K\rho^2. \ee
While in the high density limit, we obtain the ultra-relativistic EoS
\be  p= \frac{1}{3}\rho c^2, \ee
which is similar to the one describing the core of neutron stars modeled by the ideal Fermi gas.

\section{Results}
\label{sec:Results_BS}
\begin{figure}[h!]
\includegraphics[width=5.0cm,angle=270]{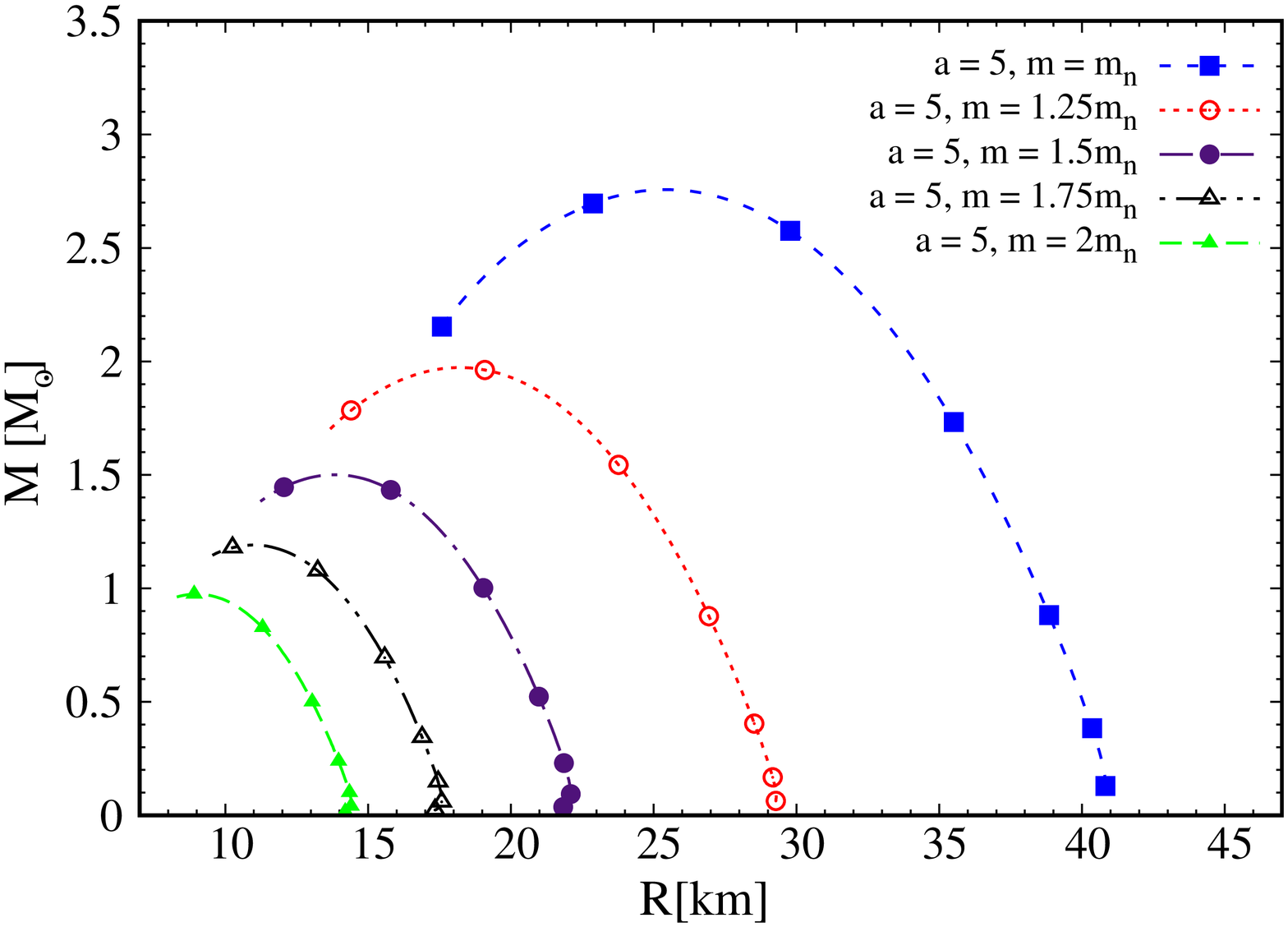}
\includegraphics[width=5.0cm,angle=270]{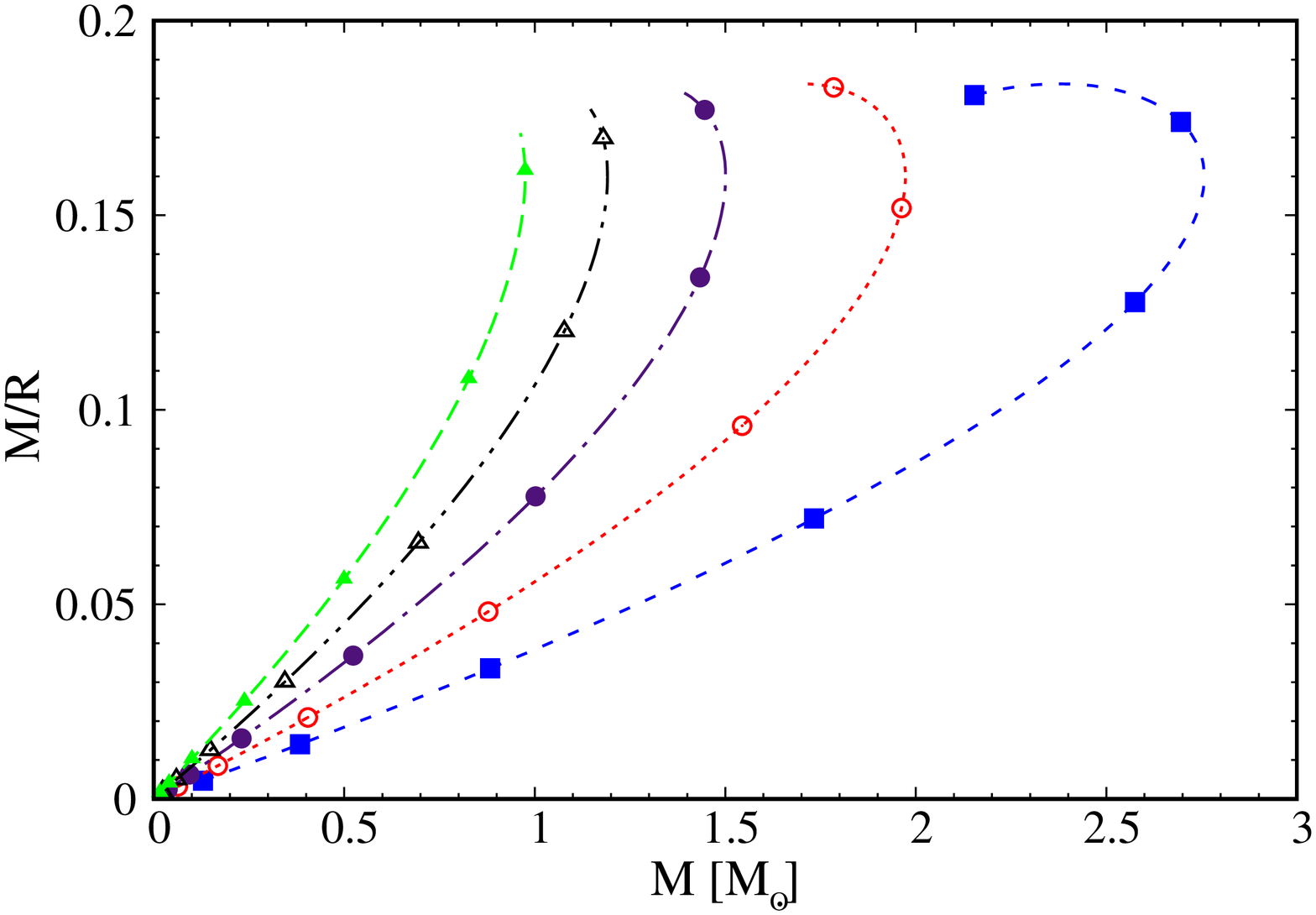}
\centering
\caption{Mass radius relation and compactness for relativistic boson star with the EoS given in
(\ref{press_BS}), we assume  $a=5 \rm{fm}$  and consider different values of the mass $m$. 
Changing these parameters it is possible to span a large range of values of mass and radius. 
From top to bottom: $m=m_n$, $m=1.25m_n$, $m=1.5m_n$, $m=1.75m_n$, and $m=2m_n$.
\label{fig:MR_EOS}}
\end{figure}

\begin{figure}[h!]
\includegraphics[width=5.0cm,angle=270]{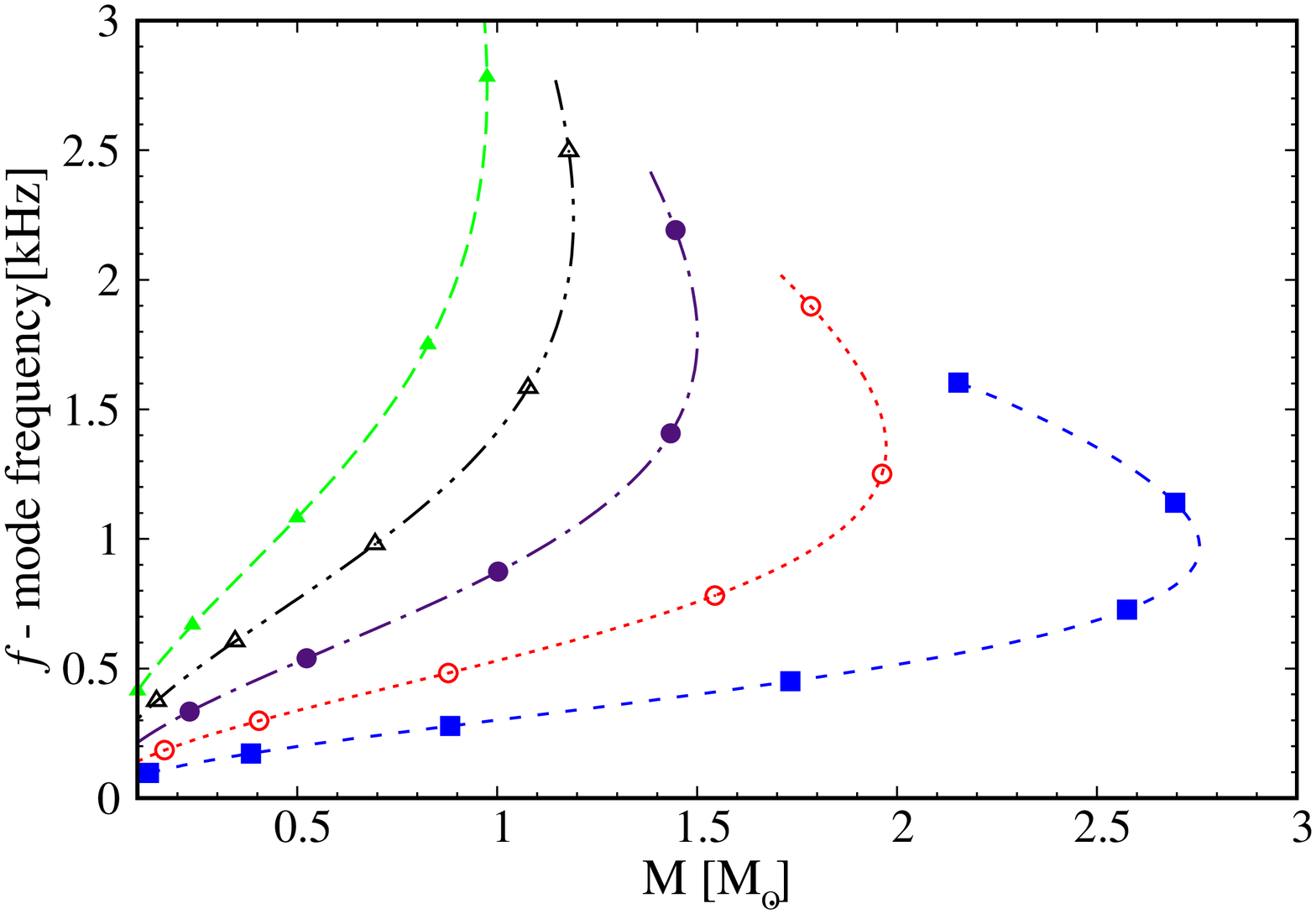}
\includegraphics[width=5.0cm,angle=270]{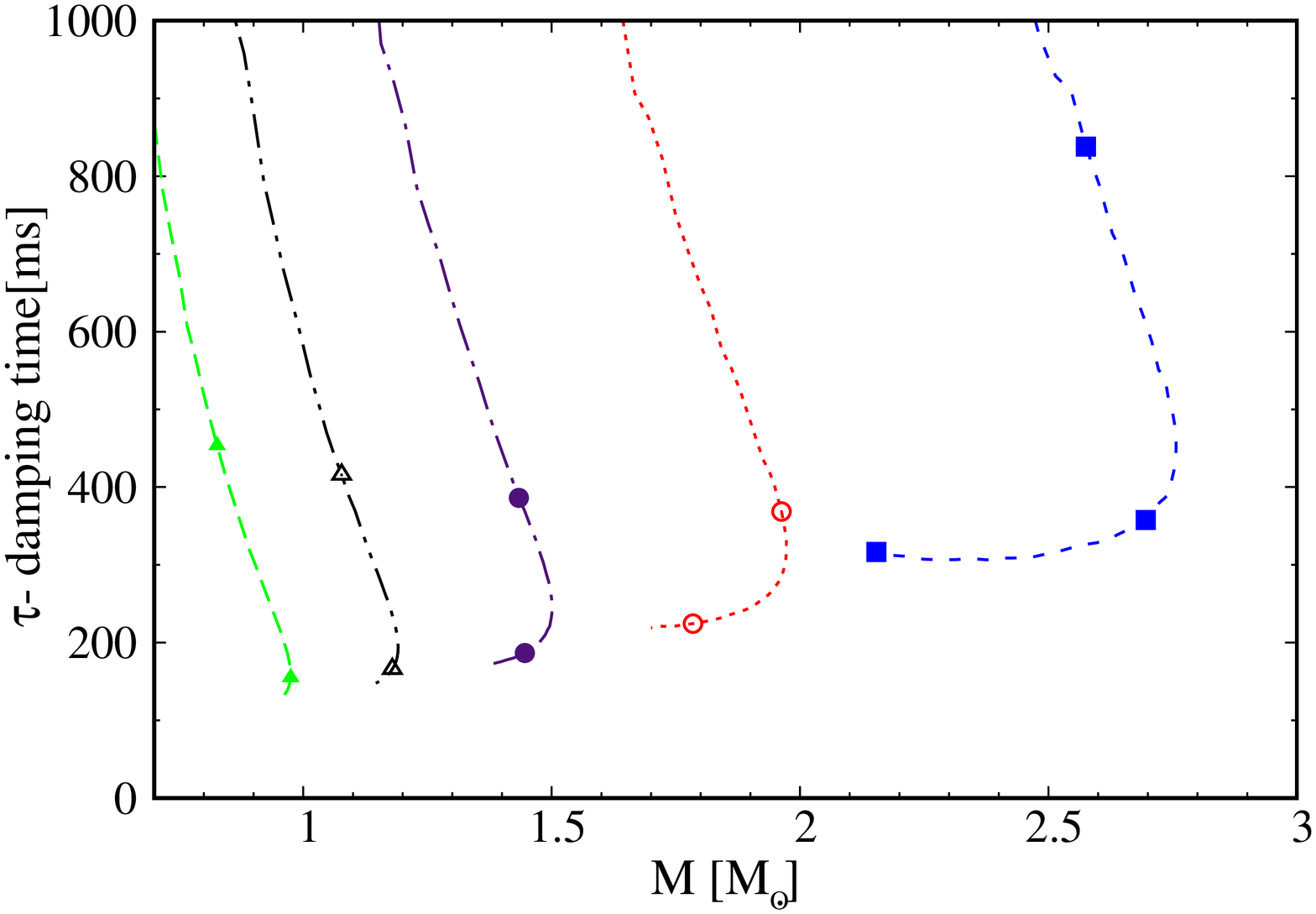}
\centering
\centering
\caption{Frequency and damping time of the fundamental mode as a function of the stellar mass, for EoS given in (\ref{press_BS})
for  $a=5 \rm{fm}$  and different values of the mass $m$.
\label{fig:frequency}}
\end{figure}

The scalar potential is symmetric under a $Z_2$ discrete transformation $\Phi \rightarrow -\Phi$.
Therefore, the lightest $Z_2$-odd component is stable and we can treat as a dark matter candidate. 
There are astronomical observations, such as the center density of DM halo, the shape of galaxy clusters and the missing satellites problem, 
would set the constraints on the self-couplings strengths and the mass of the scalar boson~\cite{Tulin:2017ara,Eby:2015hsq} within the parameter region, 
\be 0.1 \frac{\rm{cm}^2}{\rm{g}} \leq \frac{\sigma}{m} \leq 10 \frac{\rm{cm}^2}{\rm{g}}. \ee
Here $\sigma$ is the scattering cross-section among four scalars relating to $\lambda$ by
\be  \sigma=\frac{\lambda^2}{64\pi m^2}. \ee
 This interaction plays an important role in establishing an equilibrium configuration of BS. It balance in addition 
to the quantum repulsive force generated by the Heisenberg uncertainty principle with the attractive pull of gravity.
In order to make the correspondence between Bose-Einstein condensate (BEC) with short-range interactions described by the
Gross-Pitaevskii equation~\cite{Gross} and scalar fields with a $\frac{\lambda}{4}|\Phi|^4 $ interaction described by the Klein-Gordon, we set~\cite{Chavanis:2011cz}:
\be \frac{\lambda}{8\pi}\equiv\frac{a}{\lambda_c}=\frac{amc}{\hbar}. \ee
$\lambda_c=\hbar/mc$ is the Compton wavelength of $\Phi$.
In our calculations, four benchmark values of the scattering length $a$ ($a=5\rm{fm}$, $a=10\rm{fm}$, $a=15\rm{fm}$ and $a=20\rm{fm}$) and five benchmark values of mass $m$ ($1m_n$, $1.25m_n$, $1.5m_n$, $1.75m_n$ and $2m_n$, 
where $m_n$ is the neutron mass) are considered. In the main text we present the result of $a=5\rm{fm}$ and we leave the reaming three cases in the Appendix \ref{sec:appB}.
For these parameter values we calculate the  mass-radius, the compactness, the $f$-mode frequency defined as $f=\sigma/2\pi$, 
and the damping time $\tau$ of BS, see FIG.\ref{fig:MR_EOS}-\ref{fig:frequency}. 
Compactness $C$ is defined as $C\equiv M/R$, so that $C=0.5$ for a Schwarzschild BH and $C_{\rm{max}}\approx 0.18$ for a BS.


\begin{figure}[h!]
\includegraphics[width=5.0cm,angle=270]{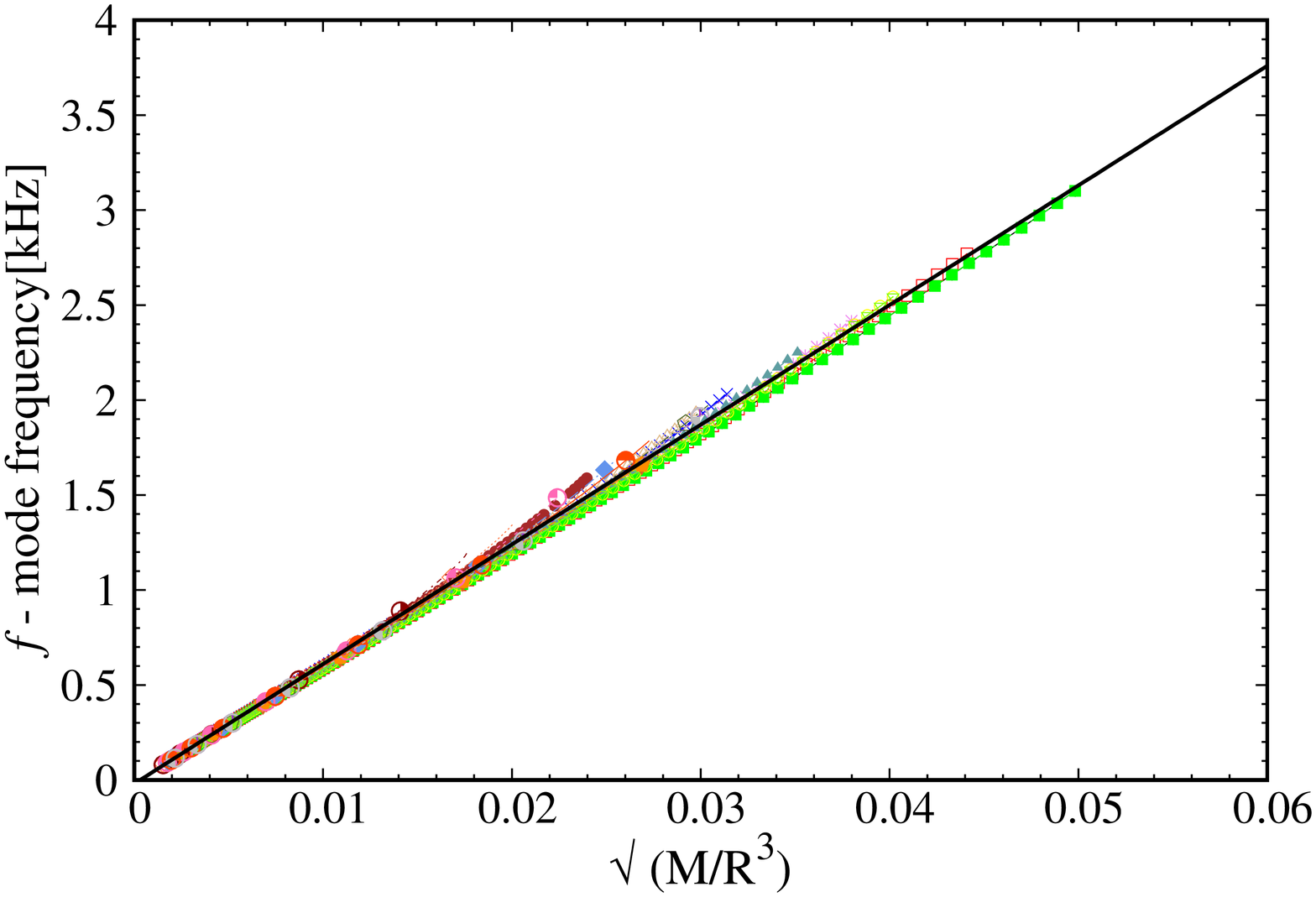}
\includegraphics[width=5.0cm,angle=270]{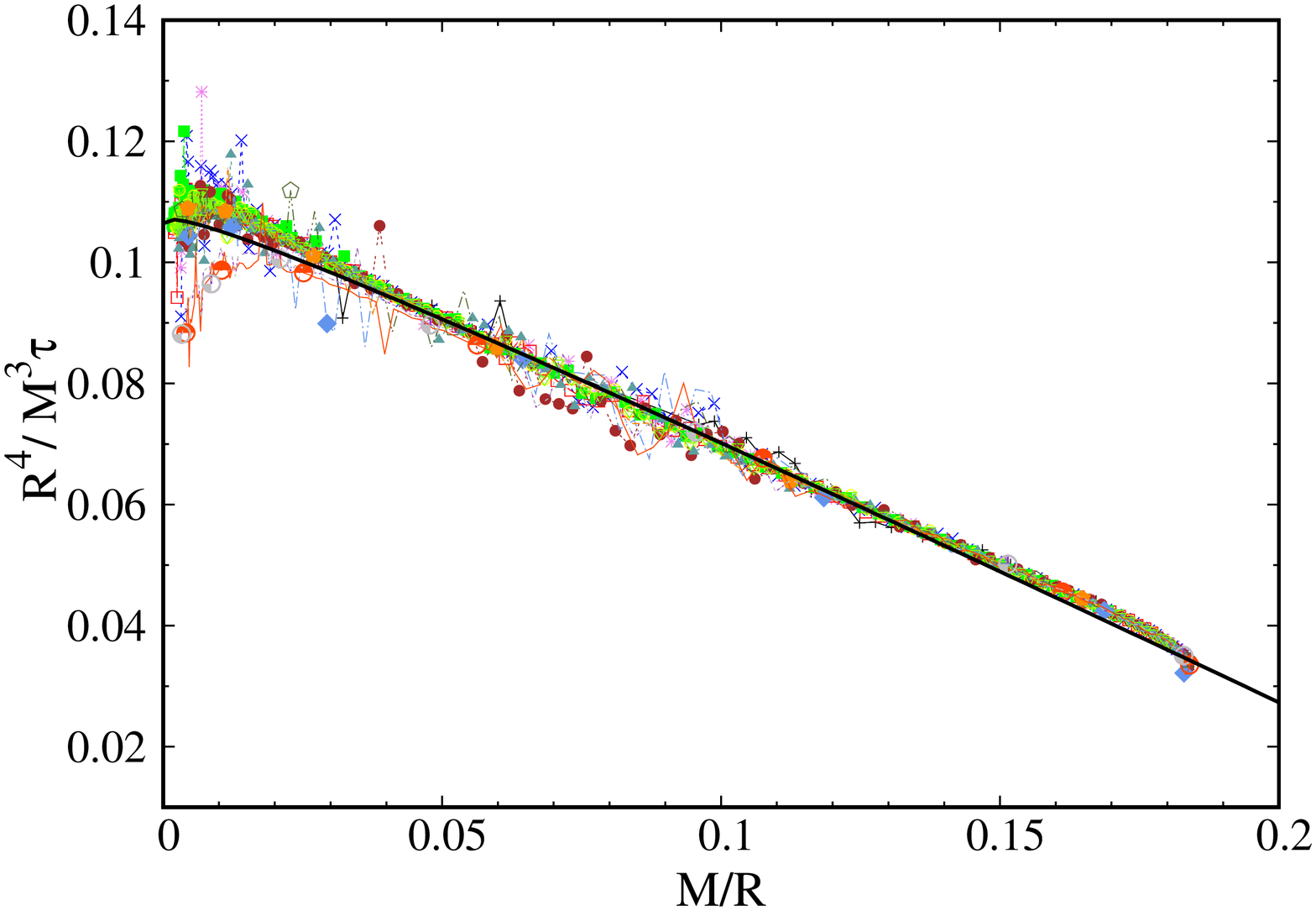}
\centering
\caption{The frequency of the fundamental mode is plotted as functions of the square root of the average stellar density,
while the normalized damping time of the $f$-modes as functions of the stellar compactness $M/R$,
for the  EoS considered in this paper with different parameter values.
\label{fig:f_universal}}
\end{figure}

\noindent Finally, we suggest two universal relations for such BS, these empirical fits were originally proposed for CSs
by Andersson and Kokkotas~\cite{Andersson:1997rn} and recently studied by Chirenti et al.~\cite{Chirenti:2015dda} for different CSs EoS.
The first is the relation between the $f$-mode frequencies and the square root of the average stellar density, 
$\sqrt{M/R^3}$, which is known to be the natural scale of the mode. We have 
\be  f=b_1+b_2\sqrt{\frac{M}{R^3}}, \ee
with $b_1=-0.0195\pm 0.0008$, $b_2=62.997 \pm 0.058$.    
And the second empirical relation on the damping time seems also to be universal for the BS:
\be  \left(\frac{M^3\tau}{R^4}\right)^{-1}=c_1+c_2\sqrt{\frac{M}{R}}+c_3\frac{M}{R} \ee
with values $c_1=0.106 \pm 0.0005$, $c_2=0.035 \pm 0.005$, and $c_3=-0.474 \pm 0.010$.
It is interesting to notice that the coefficients of this expansion do not depend on the chosen of the BSs model. 
We give all the coefficients for the various models analyzed in this paper in the table \ref{table:1}. These universal relations are shown in FIG.\ref{fig:f_universal}.

\begin{table}[ht]
\begin{center}
\caption{Set of parameters obtained for our model.}
\resizebox{0.85\textwidth}{!}{
\begin{tabular}{cccccccccccc}
\hline\hline
                                  &               & $b_{1}(\rm{kHz})$ & $b_{2}(\rm{km/kHz})$ & $c_{1}$ & $c_{2}$  & $c_{3}$  &\\ \hline
\multirow{6}{*}{$a=5[\rm{fm}]$}   & $m=1.00m_n$   & $-0.030$          & $64.93$              & $0.107$ & $0.031$  & $-0.468$ &\\ 
                                  & $m=1.25m_n$   & $-0.028$          & $63.59$              & $0.112$ & $0.003$  & $-0.430$ &\\ 
                                  & $m=1.50m_n$   & $-0.029$          & $62.79$              & $0.110$ & $0.018$  & $-0.448$ &\\ 
                                  & $m=1.75m_n$   & $-0.028$          & $62.18$              & $0.110$ & $0.018$  & $-0.451$ &\\
                                  & $m=2.00m_n$   & $-0.028$          & $61.72$              & $0.111$ & $0.011$  & $-0.444$ &\\ \hline
\multirow{6}{*}{$a=10[\rm{fm}]$}  & $m=1.00m_n$   & $-0.030$          & $66.41$              & $0.104$ & $0.038$  & $-0.473$ &\\ 
                                  & $m=1.25m_n$   & $-0.031$          & $64.98$              & $0.104$ & $0.059$  & $-0.519$ &\\ 
                                  & $m=1.50m_n$   & $-0.030$          & $63.95$              & $0.110$ & $0.009$  & $-0.429$ &\\ 
                                  & $m=1.75m_n$   & $-0.029$          & $63.14$              & $0.108$ & $0.031$  & $-0.470$ &\\
                                  & $m=2.00m_n$   & $-0.029$          & $62.56$              & $0.110$ & $0.013$  & $-0.438$ &\\ \hline
\multirow{6}{*}{$a=15[\rm{fm}]$}  & $m=1.00m_n$   & $-0.029$          & $67.26$              & $0.090$ & $0.201$  & $-0.792$ &\\ 
                                  & $m=1.25m_n$   & $-0.030$          & $65.77$              & $0.106$ & $0.037$  & $-0.484$ &\\ 
                                  & $m=1.50m_n$   & $-0.031$          & $64.72$              & $0.107$ & $0.024$  & $-0.452$ &\\ 
                                  & $m=1.75m_n$   & $-0.030$          & $63.82$              & $0.110$ & $0.015$  & $-0.442$ &\\
                                  & $m=2.00m_n$   & $-0.029$          & $63.14$              & $0.110$ & $0.014$  & $-0.463$ &\\ \hline
\multirow{6}{*}{$a=20[\rm{fm}]$}  & $m=1.00m_n$   & $-0.033$          & $68.06$              & $0.057$ & $0.356$  & $-0.981$ &\\ 
                                  & $m=1.25m_n$   & $-0.018$          & $63.17$              & $0.079$ & $0.190$  & $-0.703$ &\\ 
                                  & $m=1.50m_n$   & $-0.016$          & $61.92$              & $0.086$ & $0.154$  & $-0.646$ &\\ 
                                  & $m=1.75m_n$   & $-0.016$          & $61.30$              & $0.087$ & $0.157$  & $-0.658$ &\\
                                  & $m=2.00m_n$   & $-0.030$          & $62.29$              & $0.089$ & $0.147$  & $-0.647$ &\\                                
\hline\hline
\label{table:1}
\end{tabular}}
\end{center}
\end{table}

\section{Conclusions}
\label{sec:conclusions}
\noindent In this paper we have studied the $f$-mode of BS.
We restrict our attention to the case of massive boson stars described by a relativistic EoS given by the equation  (\ref{press_BS}), 
we choose the values of the scattering length $a$ and the boson mass $m$ in order to satisfy the self-interacting constraints on the DM. 
For these parameter values we calculate the mass, the radius, the $f$-mode frequency, the damping time and the compactness of BS.
Our results are obtained by solving the linear perturbation equations that describe the nonradial oscillations of relativistic compact stars.\\
Our results contribute to a series of empirical fits proposed in the literature that describe 
the general behavior of the $f$-mode frequency and damping time $\tau$ as functions of the stars average density and compactness~\cite{Andersson:1997rn,Chirenti:2015dda}.
We find that the relations of universality are valid for the BS not only for objects with mass comparable to the normal NS
but also for BS with masses between 3-6 solar masses compatible with those of BH. 
This could be used to identify objects of large masses that show frequencies of oscillations not compatible with those of BH.\\ 

\acknowledgments
Alessandro Parisi is grateful for the hospitality at the National Center for Theoretical Sciences (NCTS)
of Hsinchu, where part of this work was carried out, and Professor Feng-Li Lin for many helpful discussions.

\newpage

\appendix

\section{The Lindblom-Detweiler Equations}
\label{sec:appA} 
\noindent The system of the Lindblom and Detweiler equations~\cite{Lindblom:1983ps,Detweiler:1985zz} consist of four
first-order differential equations in the quantities $H_1^{\ell m}(r),K^{\ell m}(r),W^{\ell m}(r),X^{\ell m}(r)$:

\begin{eqnarray}\label{Detweiler1}
  H_1^{'\ell m} &=& -\frac{1}{r}\left[\ell+1+\frac{2M e^\lambda}{r}+4\pi  r^2e^\lambda(p-\varepsilon)\right] H_1^{\ell m}+\frac{e^\lambda}{r}[H_0^{\ell m}+K^{\ell m}-16\pi(p+\varepsilon)V^{\ell m}],\nonumber\\
  K^{'\ell m}   &=& \frac{1}{r}H_0^{\ell m}+\frac{\ell(\ell+1)}{2r}H_1^{\ell m}-\left[\frac{\ell+1}{r}-\frac{\psi'}{2}\right]K^{\ell m}-8\pi(p+\varepsilon)\frac{e^{\lambda/2}}{r}W^{\ell m}, \nonumber\\
  W^{'\ell m} &=& -\frac{\ell+1}{r}W^{\ell m}+r e^{\lambda/2}\left[\frac{e^{-\psi/2}}{(p+\varepsilon)c_s^2}X^{\ell m}-\frac{\ell(\ell+1)}{r^2}V^{\ell m}+\frac{1}{2}H_0^{\ell m}+K^{\ell m}\right], \nonumber\\
  X^{'\ell m} &=& -\frac{\ell}{r}X^{\ell m}+\frac{(p+\varepsilon)e^{\psi/2}}{2}[\left(\frac{1}{r}-\frac{\psi'}{2}\right)H_0^{\ell m}+\left(r \omega^2 e^{-\psi}+\frac{\ell(\ell+1)}{2\; r}\right)H_1^{\ell m}+\left(\frac{3}{2}\psi'-\frac{1}{r}\right)K^{\ell m}  \nonumber\\
          & & -\frac{\ell(\ell+1)}{r^2}\psi' V^{\ell m}-\frac{2}{r}\left(4\pi(p+\varepsilon)e^{\lambda/2}+\omega^2 e^{\lambda/2-\psi}-\frac{r^2}{2}\left(\frac{e^{-\lambda/2}}{r^2}\psi'\right)'\right)W^{\ell m}].
\end{eqnarray}

\noindent The remaining perturbation functions, $H_0^{\ell m}(r),V^{\ell m}(r),H_2^{\ell m}(r)$, are given by the algebraic relations: \\

\begin{eqnarray}\label{Detweiler2}
0 &=& \left[3M+\frac{1}{2}(\ell-1)(\ell+2)r+4\pi r^3 p\right] H_0^{\ell m}-8\pi r^3 e^{-\psi/2}X^{\ell m} \nonumber\\
  & & +\left[\frac{1}{2}\ell(\ell+1)(M+4\pi r^3 p)-\omega^2 r^3 e^{-(\lambda+\psi)}\right]H_1^{\ell m}\nonumber\\
  & & -\left[\frac{1}{2}(\ell-1)(\ell+2)r-\omega^2 r^3 e^{-\psi}-\frac{e^\lambda}{r}(M+4\pi r^3 p)(3M-r+4\pi r^3 p)\right]K^{\ell m}, \nonumber\\
  X^{\ell m}    &=& \omega^3(\varepsilon+p)e^{-\psi/2}V^{\ell m}-\frac{p'}{r}e^{(\psi-\lambda)/2}W^{\ell m}+\frac{1}{2}(\varepsilon+p)e^{\psi/2}H_0^{\ell m}, \nonumber\\
  H_0^{\ell m}  &=&  H_2^{\ell m}.
\end{eqnarray}

\noindent Equations (\ref{Detweiler1}) and (\ref{Detweiler2}) are solved
numerically inside the star, assuming that perturbation functions are nonsingular near the center. 
An asymptotic expansion in a power series about $r=0$  shows that the first order constraints  implies:\\

\be\label{Detweiler3} X^{\ell
m}(0)=(\varepsilon_0+p_0)e^{\psi_0/2}\left\{\left[\frac{4\pi}{3}(\varepsilon_0+3p_0)-\frac{\omega^2}{\ell}e^{-\psi_0}\right]W^{\ell
m}(0)+\frac{1}{2}K^{\ell m}(0)\right\}, \ee

\be\label{Detweiler4}  H_1^{\ell
m}(0)=\frac{1}{\ell(\ell+1)}[2\ell K^{\ell
m}(0)+16\pi(\varepsilon_0+p_0)W^{\ell m}(0)], \ee

\noindent where the constants $\varepsilon_0$, $p_0$, and $\psi_0$ appearing in these expressions are simply the first terms in the
power-series expansions for the density, pressure, and gravitational potential. On the stellar surface, $r=R$, one assumes continuity of the perturbation functions and the vanishing of the Lagrangian pressure perturbation, i.e.,

\be\label{Detweiler5} X^{\ell m}(R)=0 \ee

\noindent In the exterior, the metric perturbations are described by the Zerilli functions: 

\be  Z^{\ell m}=\frac{r^{\ell+2}}{nr+3M}(K^{\ell
m}-e^\psi H_1^{\ell m}),\ee where $n=(\ell-1)(\ell+2)/2$, which is
solution of the Zerilli equation
 \be\label{Zerilli} \frac{d^2Z^{\ell
m}}{dr_\star^2}+[\omega^2-V_Z(r)]Z^{\ell m}=0 \ee

\noindent with $r_\star\equiv r+2M \ln(r/2M-1)$ and \be V_Z\equiv
e^{-\lambda}\frac{2n^2(n+1)r^3+6n^2Mr^2+18nM^2r+18M^3}{r^3(nr+3M)^2}
\ee The transformation between $H_1^{\ell m}$,$K^{\ell m}$, and
the Zerilli function is nonsingular~\cite{1971ApJ...166..197F}.
Chandrasekhar  has proven that
the reflection and transmission coefficients obtained from the
Zerilli equation are identical to those derived from the
Regge-Wheeler equation~\cite{Regge:1957td}.

The numerical determination of modes with large imaginary parts is
difficult because the solutions of (\ref{Zerilli}) representing
outgoing and ingoing waves have the asymptotic behavior \be
Z_{\textrm{out}}\sim
e^{r_\star/\tau}\;\;\;\textit{and}\;\;\;Z_{\textrm{in}}\sim
e^{-r_\star/\tau}\ee the integrating outward
$(r\rightarrow\infty)$, give numerical errors in
$Z_{\textrm{out}}$. For this reason and in order to describe the free oscillations
of the star we must impose the outgoing wave boundary condition
\be\label{Zerilli2} Z^{\ell m}(r)\rightarrow e^{-i\omega
r_\star}\;\;\;\;\;(r\rightarrow \infty).\ee
 A solution of Eqs. (\ref{Detweiler1}) and
(\ref{Zerilli}) satisfying the boundary conditions
(\ref{Detweiler3}),(\ref{Detweiler4}),(\ref{Detweiler5}), and
(\ref{Zerilli2}) only exists for a discrete set of complex values
of the frequency $\omega=2\pi\nu+i/\tau$, the quasinormal modes of
the star.

\newpage

\section{Results for various possible parameters}
\label{sec:appB}
\noindent In this Appendix, we plot additional figures of mass-radius, compactness, frequency, and damping time of BS
for three other benchmark values of a, and five values of mass.

\begin{figure}[ht!]
\includegraphics[width=3.5cm,angle=270]{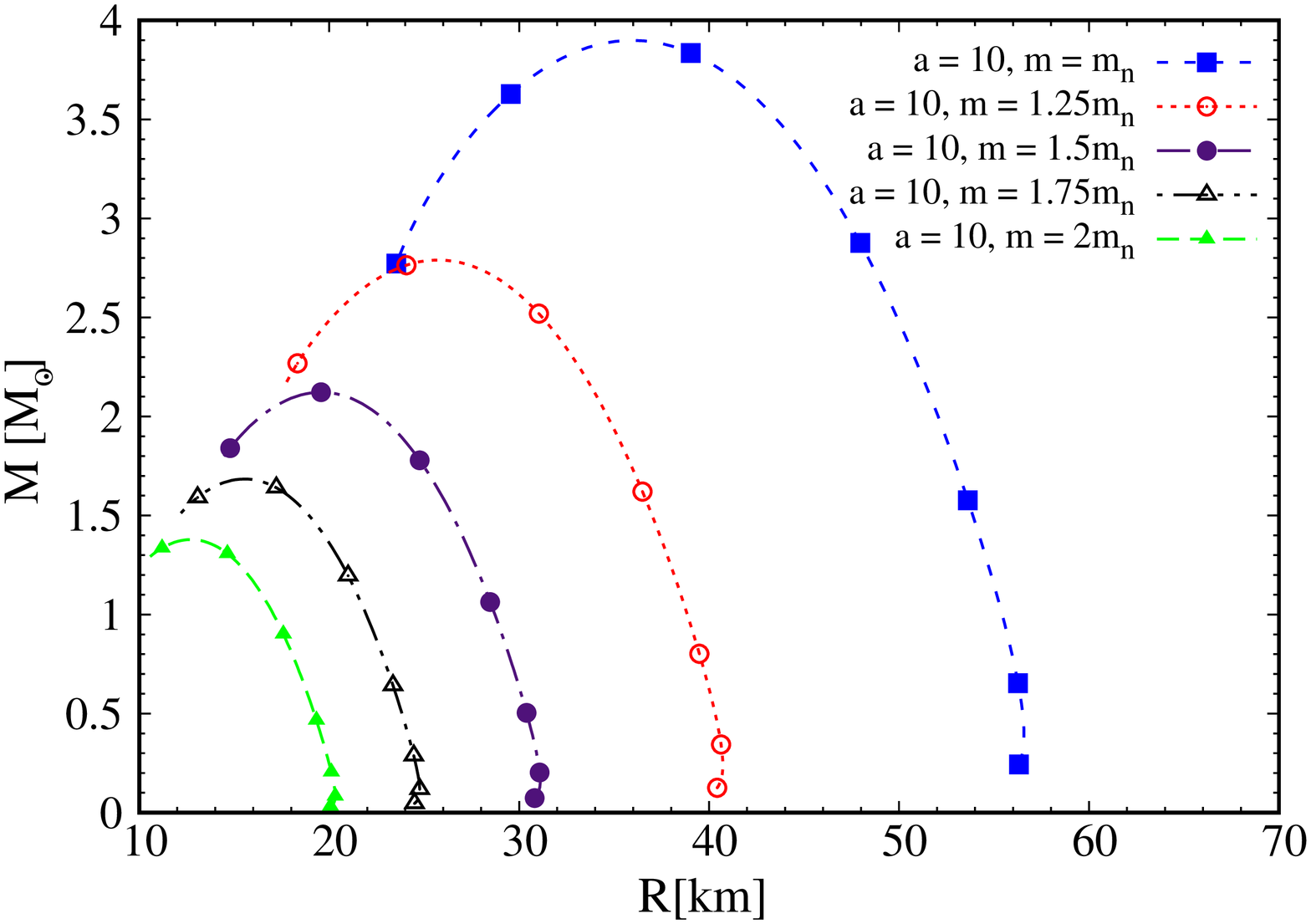}
\includegraphics[width=3.5cm,angle=270]{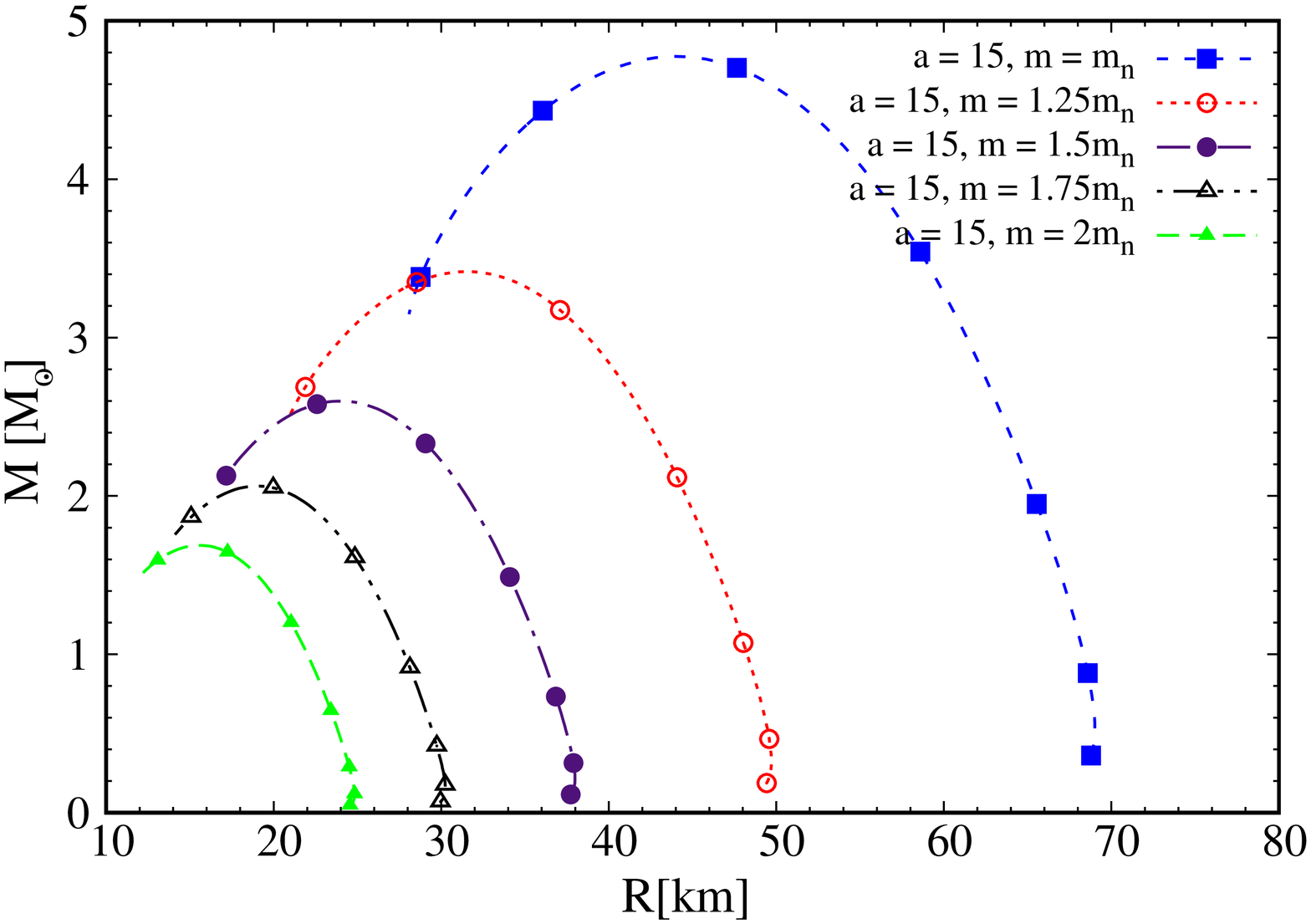}
\includegraphics[width=3.5cm,angle=270]{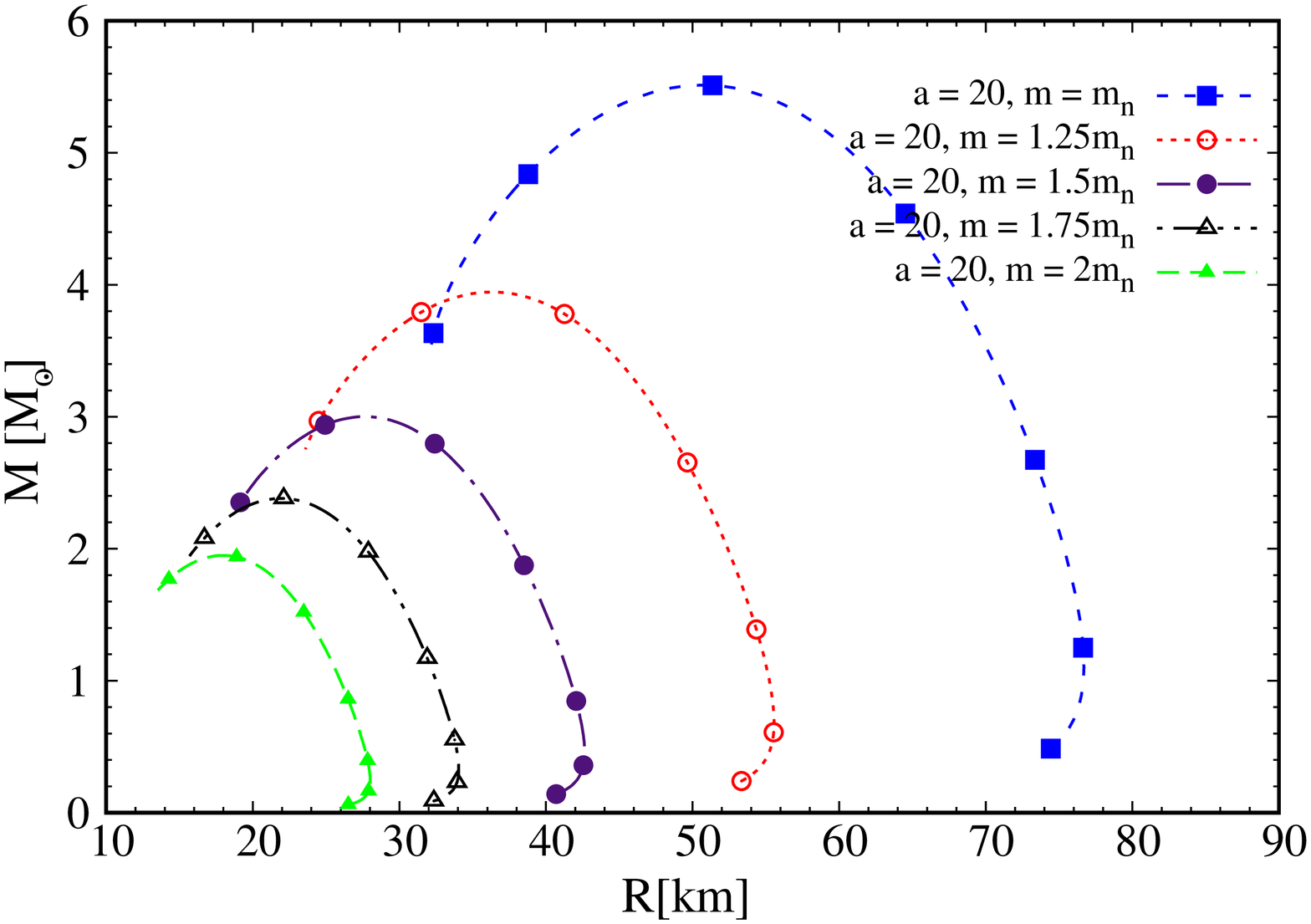}
\caption{Mass radius relation for relativistic boson star with the EoS given in
(\ref{press_BS}) for $a=10 \rm{fm}$, $a=15 \rm{fm}$,$a=20 \rm{fm}$  and different values of the mass $m$. 
Changing these parameters it is possible to span a large range of values of mass and radius. 
From top to bottom: $m=m_n$, $m=1.25m_n$, $m=1.5m_n$, $m=1.75m_n$, and $m=2m_n$.
\label{fig:MR_EOS2}}
\end{figure}

\begin{figure}[ht!]
\includegraphics[width=3.5cm,angle=270]{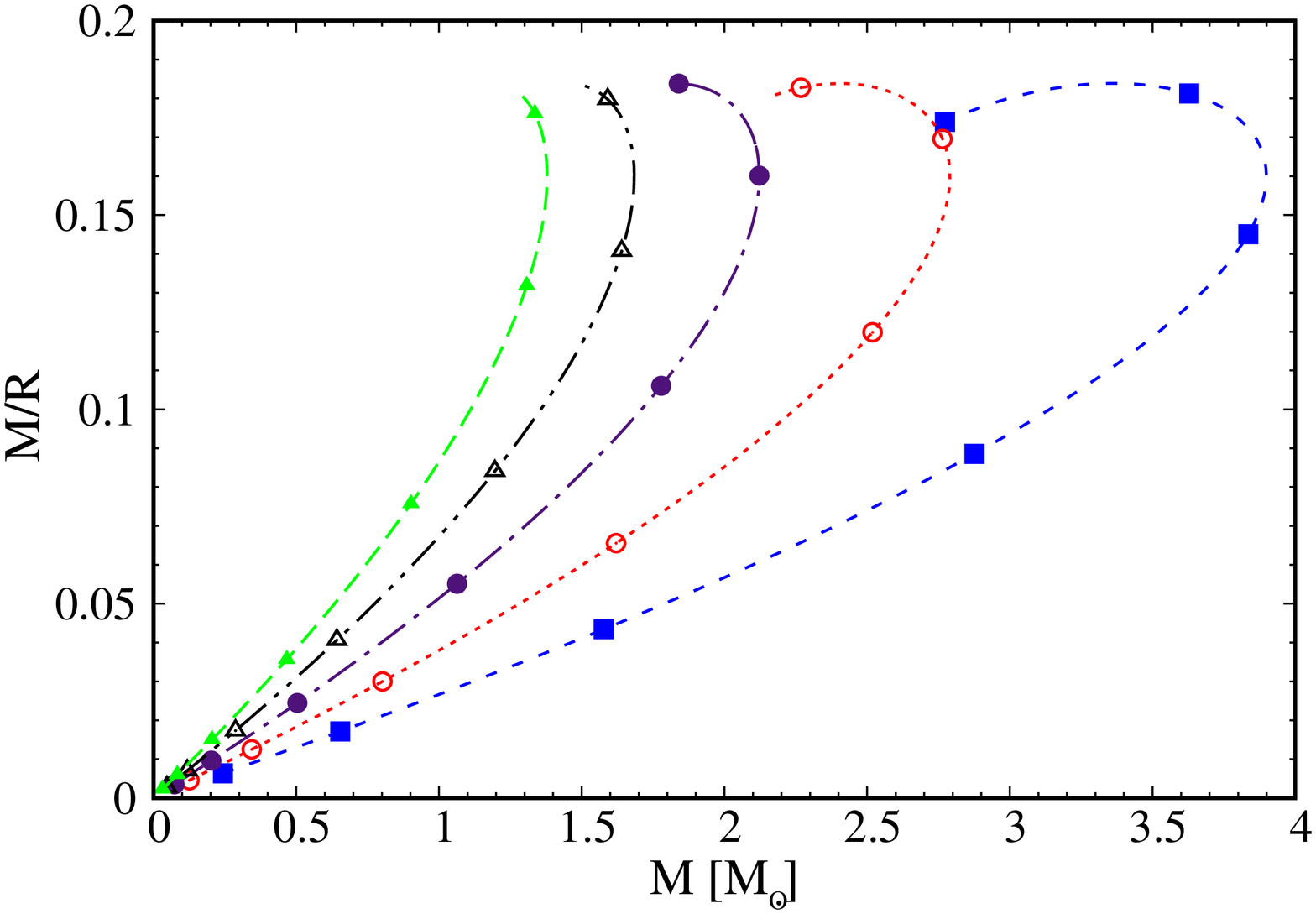}
\includegraphics[width=3.5cm,angle=270]{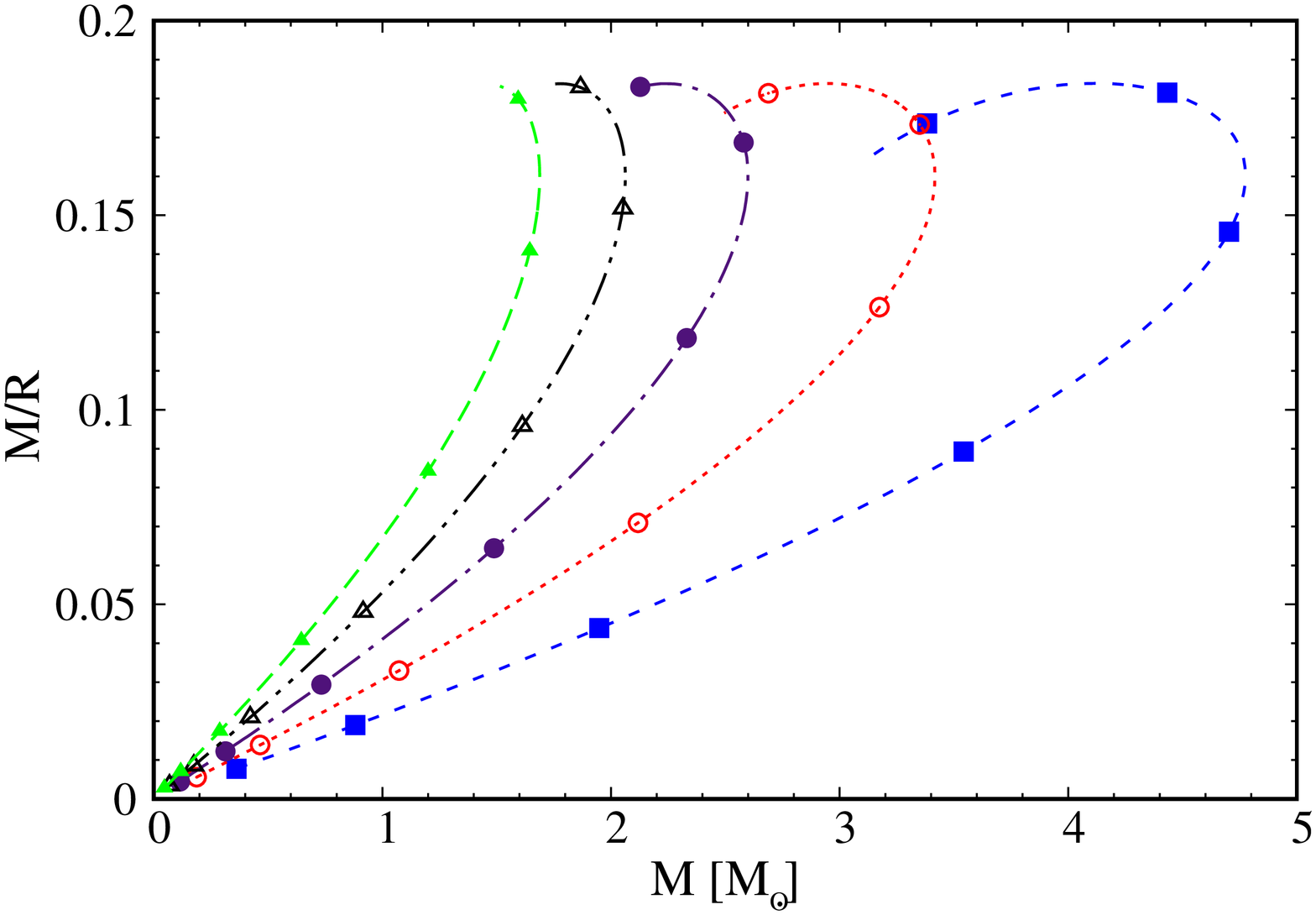}
\includegraphics[width=3.5cm,angle=270]{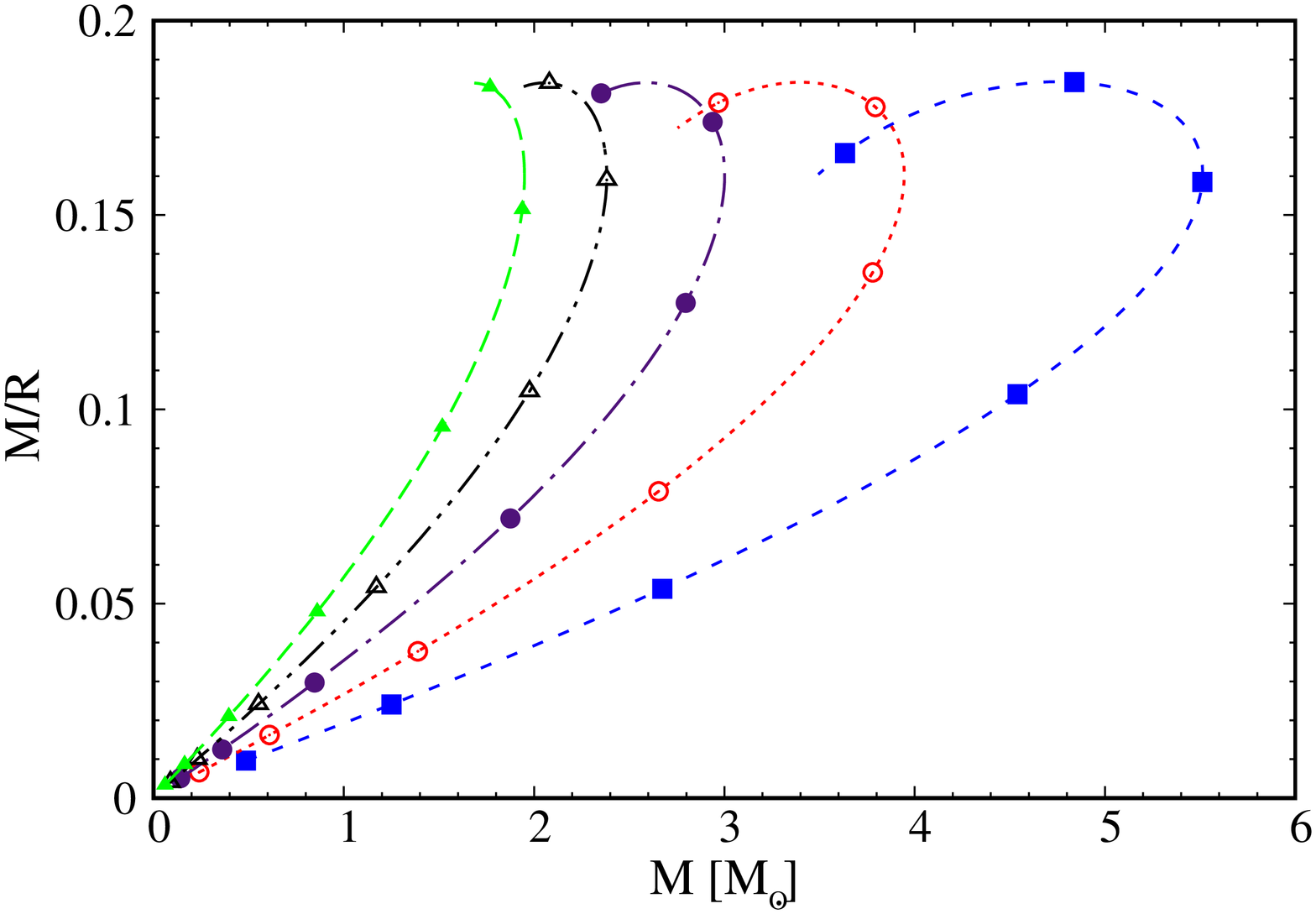}
\caption{Compactness of relativistic boson star as a function of the stellar mass, for EoS given in (\ref{press_BS}),
for three different values of the scattering length from left to right $a=10 \rm{fm}$, $a=15 \rm{fm}$, $a=20 \rm{fm}$, 
and different values of the mass: $m=m_n$, $m=1.25m_n$, $m=1.5m_n$, $m=1.75m_n$, and $m=2m_n$,
with $m = m_n$ for the blue line up to $m = 2m_n$ for the green line.
\label{fig:compactness2}}
\end{figure}

\begin{figure}[ht!]
\includegraphics[width=3.5cm,angle=270]{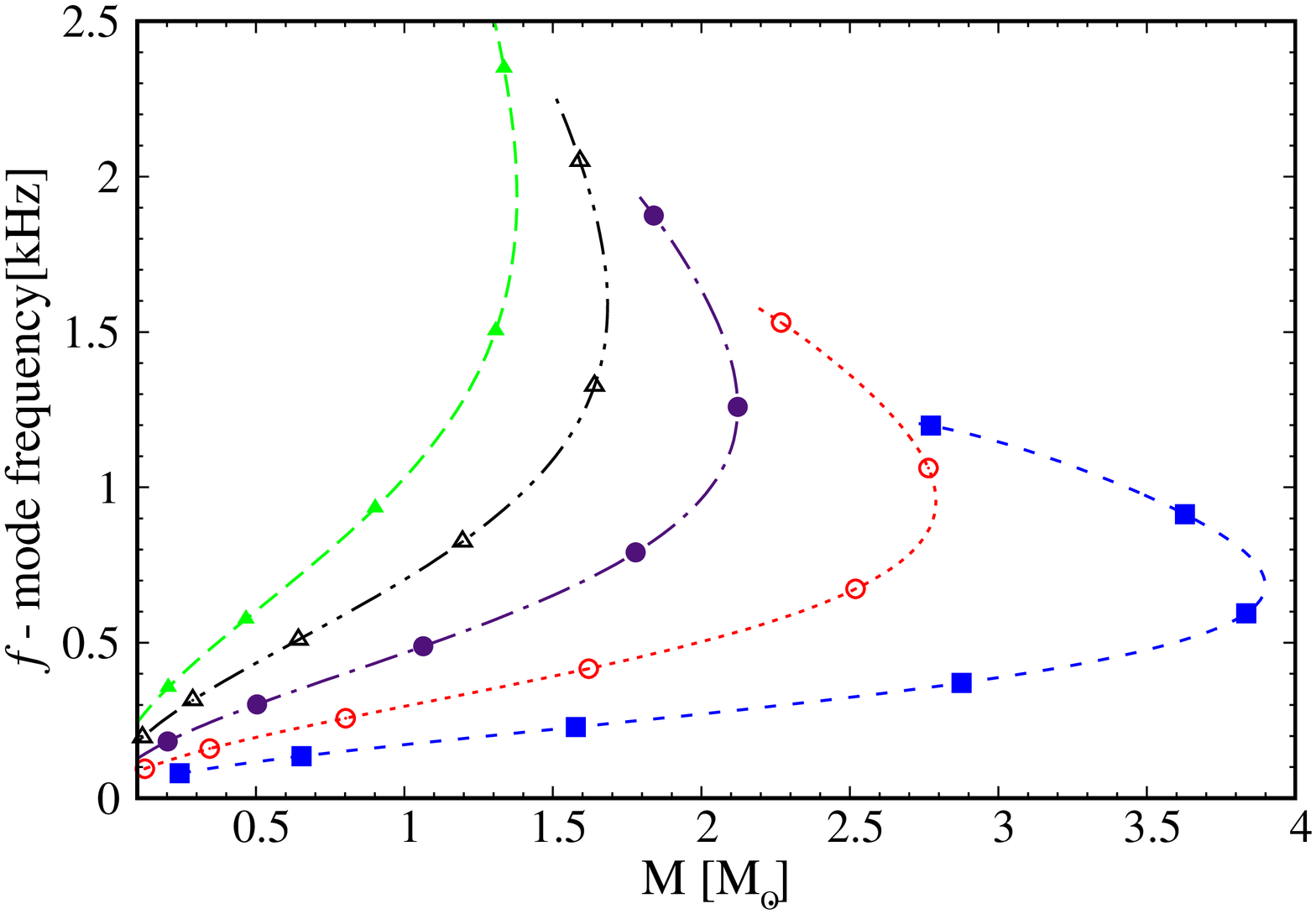}
\includegraphics[width=3.5cm,angle=270]{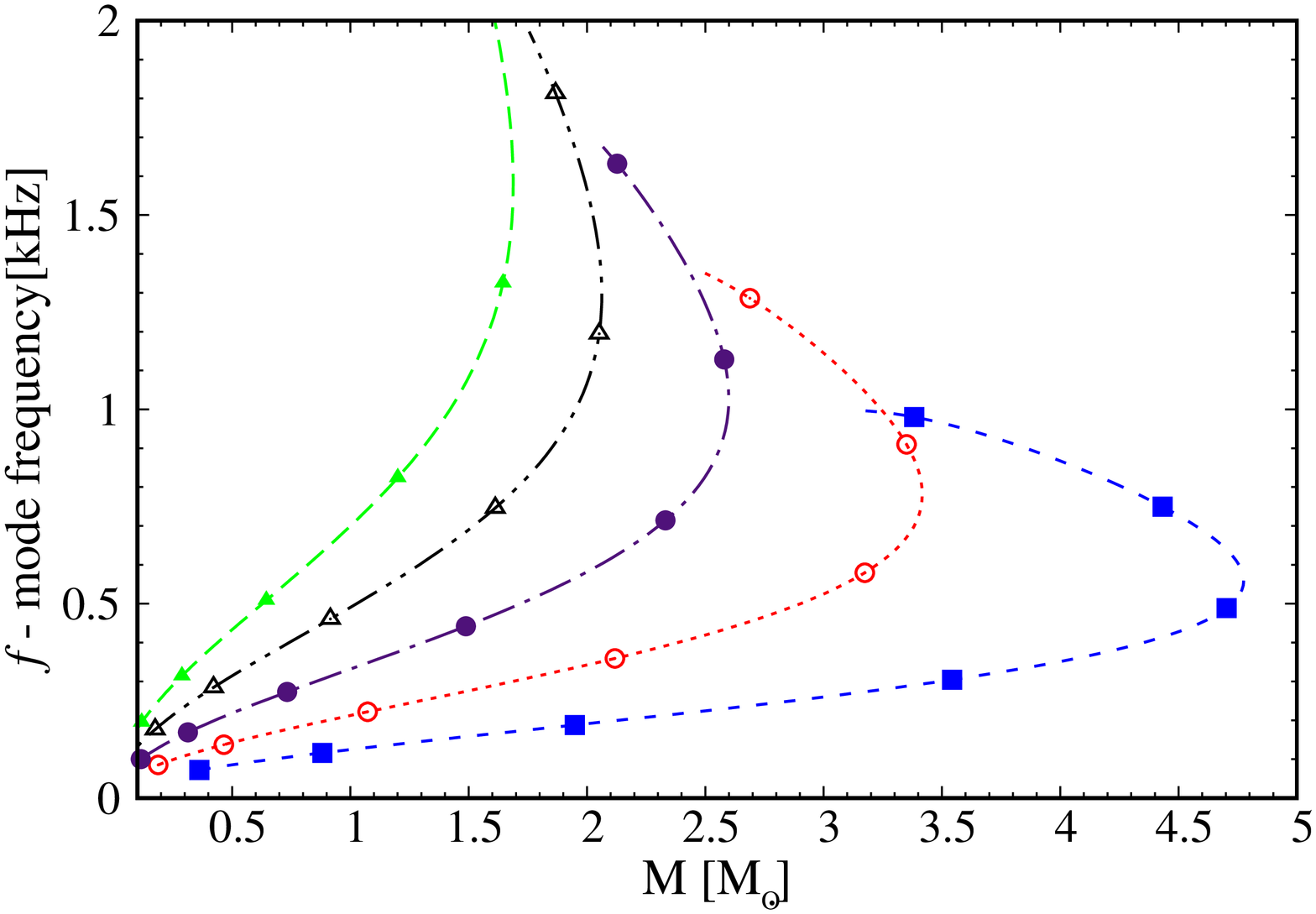}
\includegraphics[width=3.5cm,angle=270]{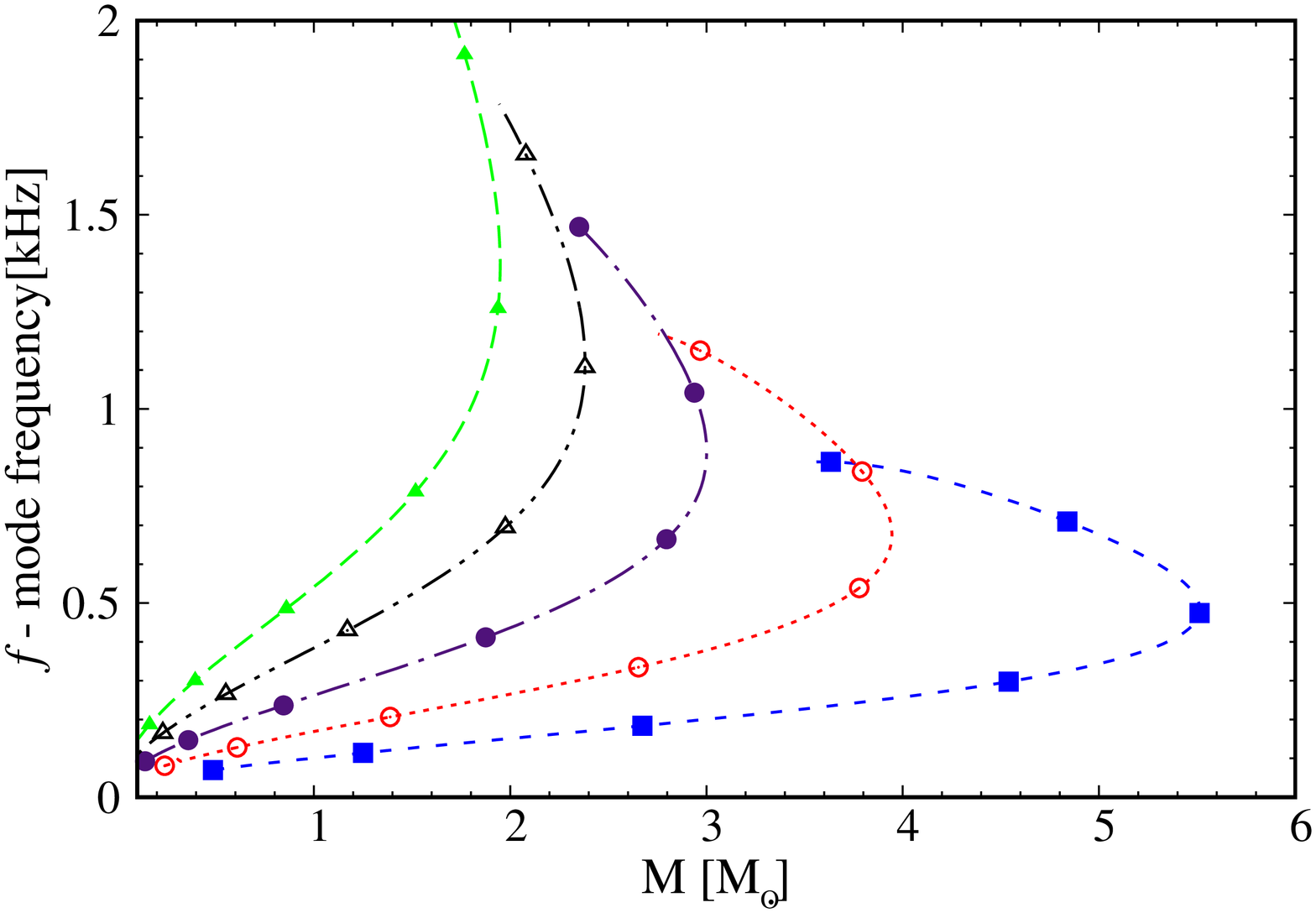}
\caption{Frequency of the fundamental mode as a function of the stellar mass, for EoS given in (\ref{press_BS}),
for three different values of the scattering length from left to right $a=10 \rm{fm}$, $a=15 \rm{fm}$, $a=20 \rm{fm}$, 
and different values of the mass: $m=m_n$, $m=1.25m_n$, $m=1.5m_n$, $m=1.75m_n$, and $m=2m_n$, 
with $m = m_n$ for the blue line up to $m = 2m_n$ for the green line.
\label{fig:frequency_damping2}}
\end{figure}

\newpage

\begin{figure}[h]
\includegraphics[width=3.5cm,angle=270]{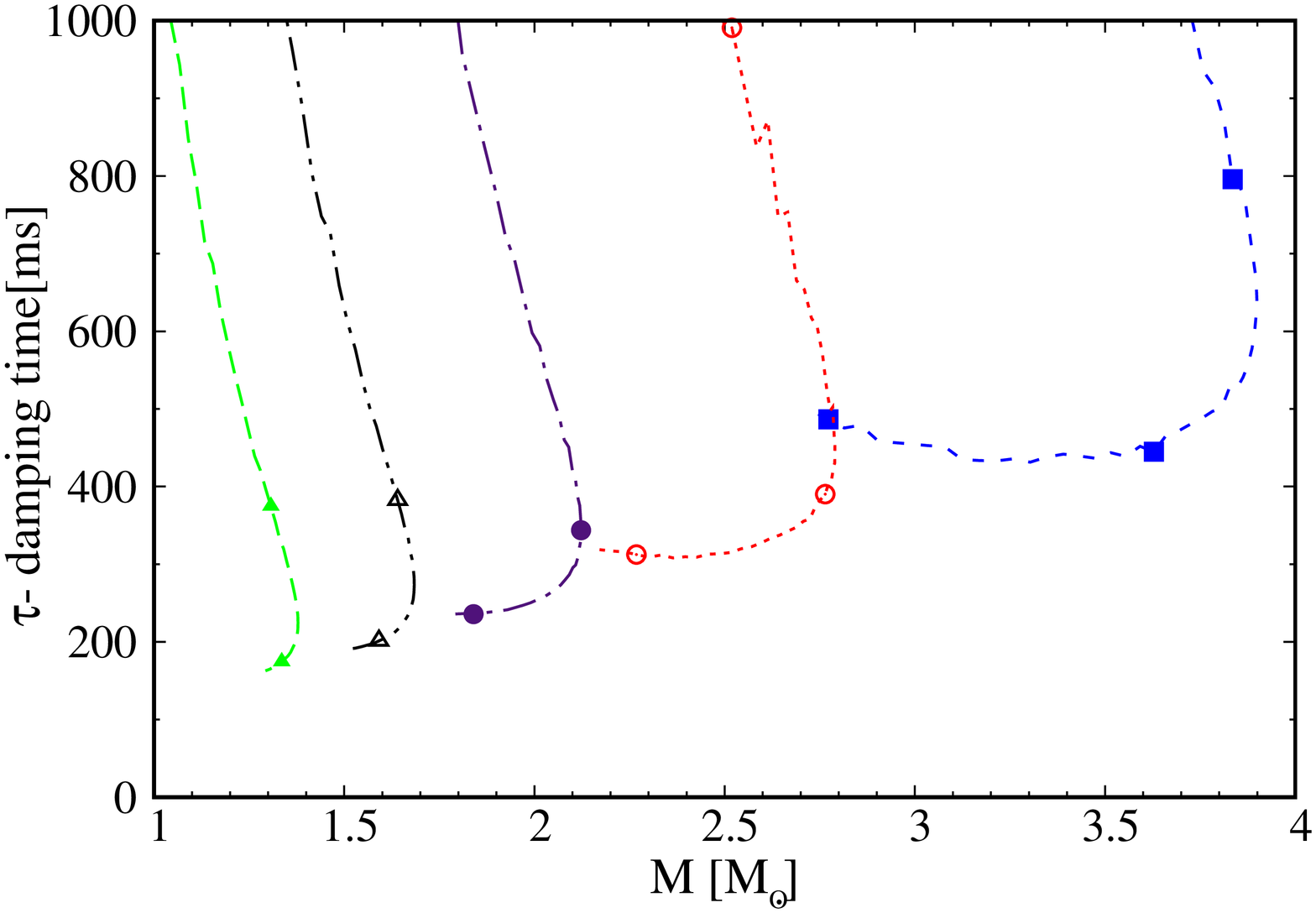}
\includegraphics[width=3.5cm,angle=270]{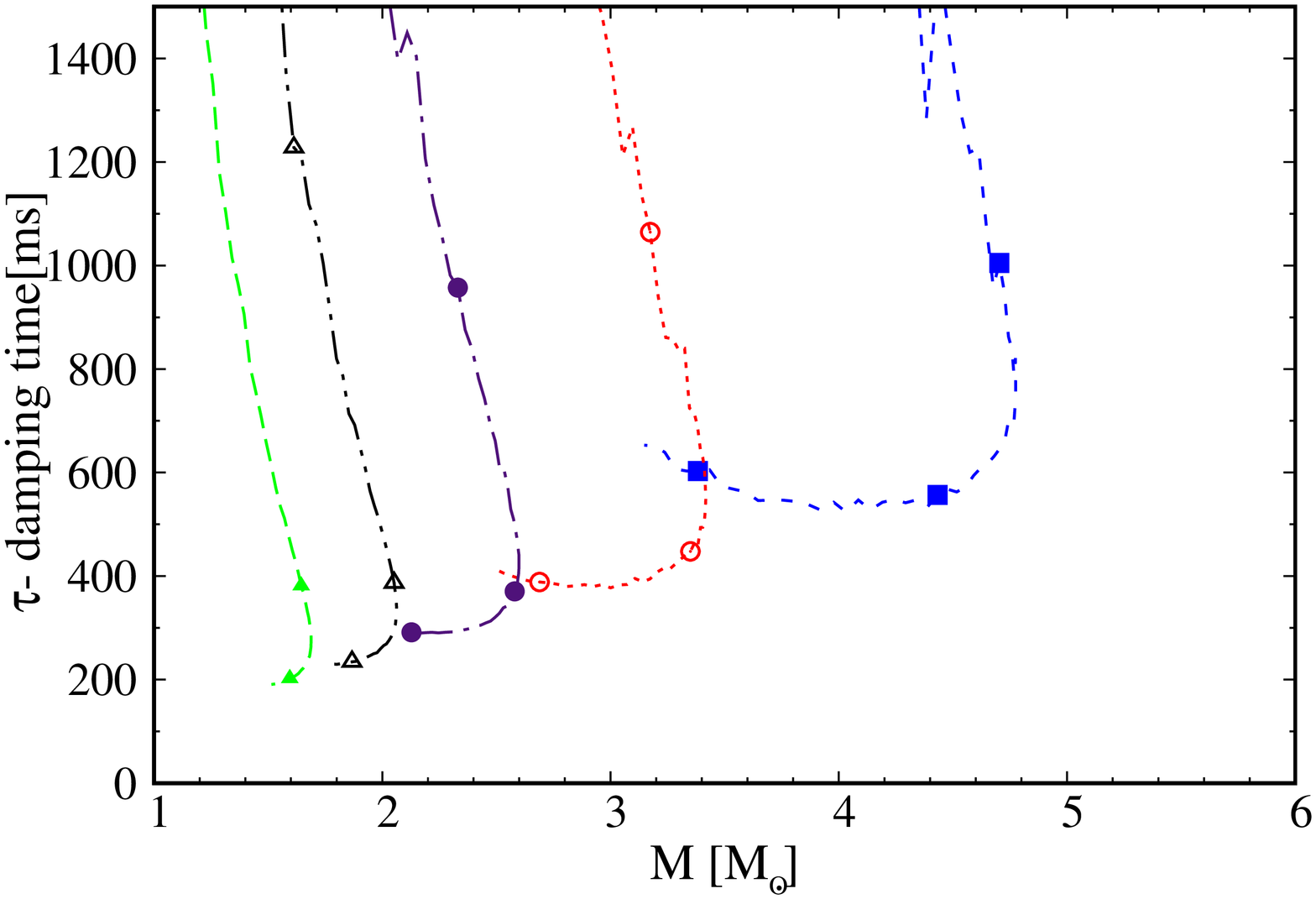}
\includegraphics[width=3.5cm,angle=270]{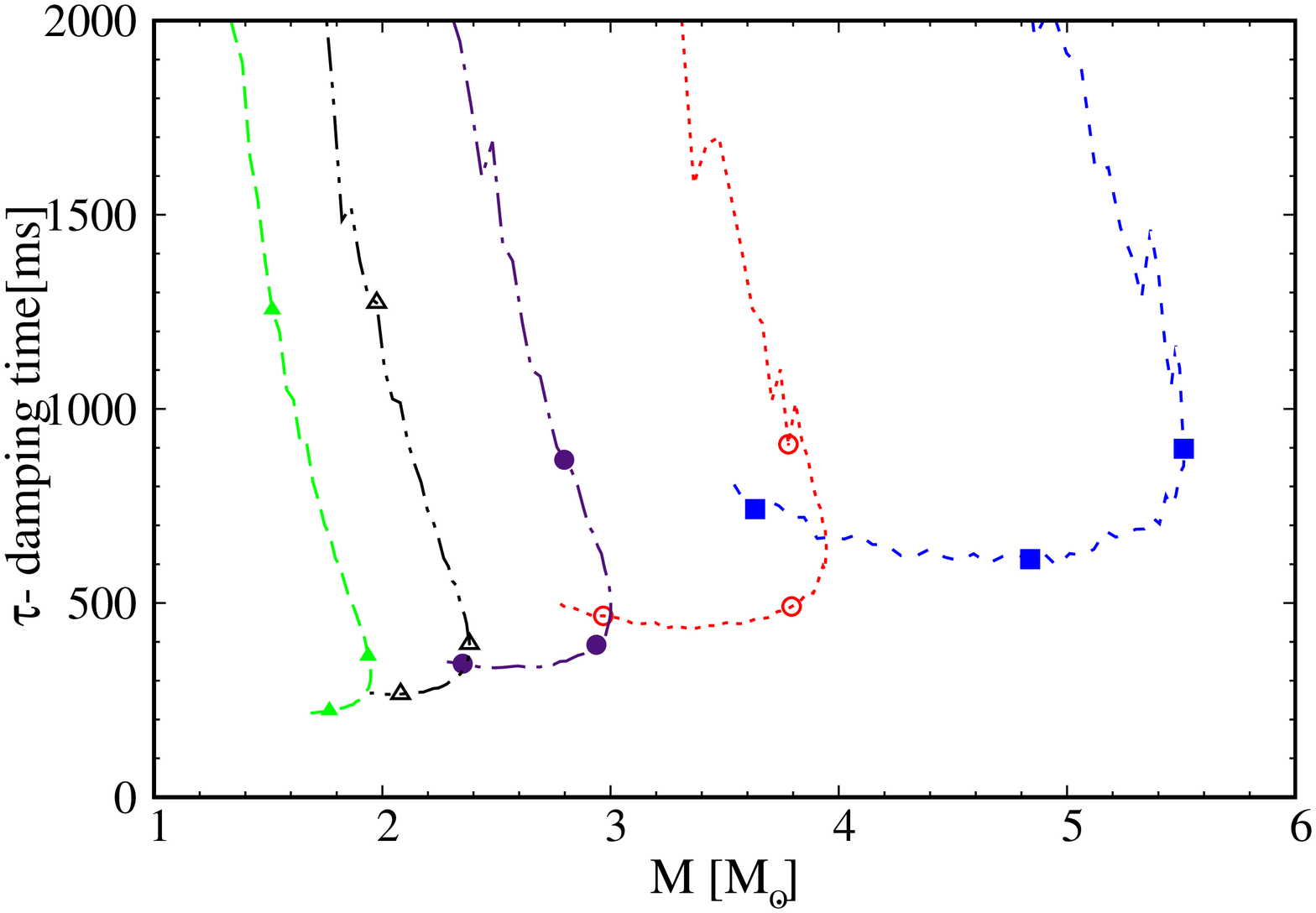}
\caption{Damping time of the fundamental mode as a function of the stellar mass, for EoS given in (\ref{press_BS}),
for three different values of the scattering length from left to right $a=10 \rm{fm}$, $a=15 \rm{fm}$, $a=20 \rm{fm}$, 
and different values of the mass: $m=m_n$, $m=1.25m_n$, $m=1.5m_n$, $m=1.75m_n$, and $m=2m_n$,
with $m = m_n$ for the blue line up to $m = 2m_n$ for the green line.
\label{fig:frequency_damping2}}
\end{figure}

\bibliographystyle{h}

\begin{thebibliography}{99}

\bibitem{Abbott:2016blz} B.~P.~Abbott {\it et al.} [LIGO Scientific and Virgo Collaborations], Phys.\ Rev.\ Lett.\  {\bf 116} (2016) no.6,  061102
  
\bibitem{TheLIGOScientific:2017qsa} B.~P.~Abbott {\it et al.} [LIGO Scientific and Virgo Collaborations], Phys.\ Rev.\ Lett.\  {\bf 119}, no. 16, 161101 (2017).

\bibitem{Ruiz:2017due} M.~Ruiz, S.~L.~Shapiro and A.~Tsokaros, Phys.\ Rev.\ D {\bf 97}, no. 2, 021501 (2018).
  
\bibitem{Annala:2017llu} E.~Annala, T.~Gorda, A.~Kurkela and A.~Vuorinen, Phys.\ Rev.\ Lett.\  {\bf 120}, no. 17, 172703 (2018).
  
\bibitem{Bauswein:2017vtn} A.~Bauswein, O.~Just, H.~T.~Janka and N.~Stergioulas, Astrophys.\ J.\  {\bf 850}, no. 2, L34 (2017).
  
\bibitem{Margalit:2017dij} B.~Margalit and B.~D.~Metzger, Astrophys.\ J.\  {\bf 850}, no. 2, L19 (2017).

\bibitem{Fattoyev:2017jql} F.~J.~Fattoyev, J.~Piekarewicz and C.~J.~Horowitz, Phys.\ Rev.\ Lett.\  {\bf 120}, no. 17, 172702 (2018).

\bibitem{Rezzolla:2017aly} L.~Rezzolla, E.~R.~Most and L.~R.~Weih, Astrophys.\ J.\  {\bf 852}, no. 2, L25 (2018).
  
\bibitem{Drago:2017bnf} A.~Drago and G.~Pagliara, Astrophys.\ J.\  {\bf 852}, no. 2, L32 (2018).

\bibitem{Nandi:2017rhy} R.~Nandi and P.~Char, Astrophys.\ J.\  {\bf 857}, no. 1, 12 (2018).

\bibitem{Wheeler:1955zz} J.~A.~Wheeler, Phys.\ Rev.\  {\bf 97}, 511 (1955).

\bibitem{Kaup:1968zz} D.~J.~Kaup, Phys.\ Rev.\  {\bf 172}, 1331 (1968).
  
\bibitem{Ruffini:1969qy} R.~Ruffini and S.~Bonazzola, Phys.\ Rev.\  {\bf 187}, 1767 (1969).

\bibitem{Liddle:1993ha} A.~R.~Liddle and M.~S.~Madsen, Int.\ J.\ Mod.\ Phys.\ D {\bf 1}, 101 (1992).

\bibitem{Aad:2012tfa} G.~Aad {\it et al.} [ATLAS Collaboration], Phys.\ Lett.\ B {\bf 716} (2012) 1

\bibitem{Chatrchyan:2012xdj} S.~Chatrchyan {\it et al.} [CMS Collaboration],  Phys.\ Lett.\ B {\bf 716} (2012) 30

\bibitem{Seidel:1993zk} E.~Seidel and W.~M.~Suen, Phys.\ Rev.\ Lett.\  {\bf 72} (1994) 2516  
  
\bibitem{Liebling:2012fv} S.~L.~Liebling and C.~Palenzuela, Living Rev.\ Rel.\  {\bf 15}, 6 (2012).

\bibitem{Jetzer:1991jr} P.~Jetzer, Phys.\ Rept.\  {\bf 220}, 163 (1992).

\bibitem{Schunck:2003kk} F.~E.~Schunck and E.~W.~Mielke, Class.\ Quant.\ Grav.\  {\bf 20}, R301 (2003).

\bibitem{Lee:1988av} T.~D.~Lee and Y.~Pang, Nucl.\ Phys.\ B {\bf 315}, 477 (1989).
  
\bibitem{Jetzer:1988vr} P.~Jetzer, Nucl.\ Phys.\ B {\bf 316}, 411 (1989).

\bibitem{Gleiser:1988rq} M.~Gleiser, Phys.\ Rev.\ D {\bf 38}, 2376 (1988).

\bibitem{RindlerDaller:2011kx} T.~Rindler-Daller and P.~R.~Shapiro, Mon.\ Not.\ Roy.\ Astron.\ Soc.\  {\bf 422}, 135 (2012). 

\bibitem{UrenaLopez:2010ur} L.~A.~Urena-Lopez and A.~Bernal, Phys.\ Rev.\ D {\bf 82}, 123535 (2010).
  
\bibitem{Yuan:2004sv} Y.~F.~Yuan, R.~Narayan and M.~J.~Rees, Astrophys.\ J.\  {\bf 606}, 1112 (2004). 

\bibitem{Torres:2002td} D.~F.~Torres, Nucl.\ Phys.\ B {\bf 626} (2002) 377
  
\bibitem{Torres:2000dw} D.~F.~Torres, S.~Capozziello and G.~Lambiase, Phys.\ Rev.\ D {\bf 62}, 104012 (2000).

\bibitem{Olivares:2018abq} H.~Olivares {\it et al.}, arXiv:1809.08682 [gr-qc].

\bibitem{Palenzuela:2007dm} C.~Palenzuela, L.~Lehner and S.~L.~Liebling, Phys.\ Rev.\ D {\bf 77}, 044036 (2008)
  
\bibitem{Palenzuela:2017kcg} C.~Palenzuela, P.~Pani, M.~Bezares, V.~Cardoso, L.~Lehner and S.~Liebling, Phys.\ Rev.\ D {\bf 96}, no. 10, 104058 (2017).

\bibitem{Ferrell:1989kz} R.~Ferrell and M.~Gleiser, Phys.\ Rev.\ D {\bf 40}, 2524 (1989).

\bibitem{Yoshida:1994xi} S.~Yoshida, Y.~Eriguchi and T.~Futamase, Phys.\ Rev.\ D {\bf 50} (1994) 6235.

\bibitem{Balakrishna:1997ej} J.~Balakrishna, E.~Seidel and W.~M.~Suen, Phys.\ Rev.\ D {\bf 58} (1998) 104004
  
\bibitem{Macedo:2013jja} C.F.B.~Macedo, P.Pani, V.Cardoso and L.C.B.Crispino, Phys.\ Rev.\ D {\bf 88} (2013) no.6,  064046

\bibitem{Kling:2017mif} F.~Kling and A.~Rajaraman, Phys.\ Rev.\ D {\bf 96} (2017) no.4,  044039
 
\bibitem{Kling:2017hjm} F.~Kling and A.~Rajaraman, Phys.\ Rev.\ D {\bf 97} (2018) no.6,  063012
  
\bibitem{Sennett:2017etc} N.~Sennett, T.~Hinderer, J.~Steinhoff, A.~Buonanno and S.~Ossokine, Phys.\ Rev.\ D {\bf 96} (2017) no.2,  024002

\bibitem{Mendes:2016vdr} R.~F.~P.~Mendes and H.~Yang, Class.\ Quant.\ Grav.\  {\bf 34} (2017) no.18,  185001

\bibitem{Flores:2017hpb} C.~V.~Flores and G.~Lugones, Phys.\ Rev.\ C {\bf 95} (2017) no.2,  025808.
  
\bibitem{Flores:2018pnn} C.~V.~Flores and G.~Lugones, JCAP {\bf 1808} (2018) no.08,  046.

\bibitem{Parisi:2017kgx} A.~Parisi and R.~Sturani, Phys.\ Rev.\ D {\bf 97} (2018) no.4,  043015.

\bibitem{Yang:2018bzx} H.~Yang, W.~E.~East, V.~Paschalidis, F.~Pretorius and R.~F.~P.~Mendes, Phys.\ Rev.\ D {\bf 98} (2018) no.4,  044007
  
\bibitem{1967ApJ...149..591T} Thorne, K.~S., \& Campolattaro, A.\ 1967, \apj, 149, 591

\bibitem{1970ApJ...159..847C} Campolattaro, A., \& Thorne, K.~S.\ 1970, \apj, 159, 847

\bibitem{Lindblom:1983ps} L.~Lindblom and S.~L.~Detweiler, Astrophys.\ J.\ Suppl.\  {\bf 53}, 73 (1983).
  
\bibitem{Detweiler:1985zz} S.~L.~Detweiler and L.~Lindblom,  Astrophys.\ J.\  {\bf 292}, 12 (1985).

\bibitem{Regge:1957td}  T.~Regge and J.~A.~Wheeler,  Phys.\ Rev.\  {\bf 108}, 1063 (1957).

\bibitem{Burgio:2011qe} G.~F.~Burgio, V.~Ferrari, L.~Gualtieri and H.-J.~Schulze, Phys.\ Rev.\ D {\bf 84} (2011) 044017

\bibitem{Bilic:2000ef} N.~Bili\'c and H.~Nikoli\'c, Nucl.\ Phys.\ B {\bf 590} (2000) 575 

\bibitem{Latifah:2014ima} S.~Latifah, A.~Sulaksono and T.~Mart, Phys.\ Rev.\ D {\bf 90} (2014) no.12,  127501

\bibitem{Gross} L.Pitaevskii and S.Stringari, Bose-Einstein Condensation, Press, Oxford, 2003.
  
\bibitem{Shapiro:1983du} S.L.Shapiro and S.A.Teukolsky  {\em {Black Holes, White Dwarfs, and Neutron Stars }}, John Wiley \& Sons, New York, (1983).

\bibitem{Gleiser:1988ih}M.~Gleiser and R.~Watkins, Nucl.\ Phys.\ B {\bf 319}, 733 (1989).
 
\bibitem{Colpi:1986ye}  M.~Colpi, S.~L.~Shapiro and I.~Wasserman, Phys.\ Rev.\ Lett.\  {\bf 57}, 2485 (1986).

\bibitem{Chavanis:2011cz} P.~H.~Chavanis and T.~Harko, Phys.\ Rev.\ D {\bf 86} (2012) 064011

\bibitem{AmaroSeoane:2010qx} P.~Amaro-Seoane, J.~Barranco, A.~Bernal and L.~Rezzolla, JCAP {\bf 1011} (2010) 002
  
\bibitem{Maselli:2017vfi} A.~Maselli, P.~Pnigouras, N.~G.~Nielsen, C.~Kouvaris and K.~D.~Kokkotas, Phys.\ Rev.\ D {\bf 96} (2017) no.2,  023005
  
\bibitem{Tulin:2017ara} S.~Tulin and H.~B.~Yu, Phys.\ Rept.\  {\bf 730}, 1 (2018)
  
\bibitem{Eby:2015hsq} J.~Eby, C.~Kouvaris, N.~G.~Nielsen and L.~C.~R.~Wijewardhana, JHEP {\bf 1602}, 028 (2016).

\bibitem{Andersson:1997rn} N.~Andersson and K.~D.~Kokkotas, Mon.\ Not.\ Roy.\ Astron.\ Soc.\  {\bf 299}, 1059 (1998)

\bibitem{Chirenti:2015dda} C.~Chirenti, G.~H.~de Souza and W.~Kastaun, Phys.\ Rev.\ D {\bf 91} (2015) no.4,  044034 
  
\bibitem{1971ApJ...166..197F} E.D.Fackerell, \apj, 166: 197-206 (1971).

\end{thebibliography}

\end{document}